\documentclass[useAMS,usenatbib]{mn2e}
\usepackage{amssymb,amsmath}
\usepackage{graphicx}
\usepackage{float}
\usepackage{afterpage}
\usepackage[toc,page]{appendix}
\usepackage{color}
\usepackage{multirow}
\usepackage{url}

\usepackage{comment}

\title[{Using  neural networks to estimate redshift distributions. An  application to CFHTLenS}]{Using  neural networks to estimate redshift distributions. An  application to CFHTLenS}
\author[Christopher Bonnett]{Christopher Bonnett$^{1}$\thanks{E-mail:
c.bonnett@gmail.com}  \\
$^{1}$Institut de Ciencies de lÕEspai, CSIC/IEEC, F. de Ciencies, Torre C5 par-2, Barcelona 08193, Spain\\
}
\begin{document}

\date{4 December 2013}

\pagerange{\pageref{firstpage}--\pageref{lastpage}} \pubyear{2013}

\maketitle

\label{firstpage}

\begin{abstract}
We present a novel way of using neural networks  (NN) to estimate the redshift distribution of a galaxy sample.
We are able to obtain a probability density function (PDF) for each galaxy using a classification neural network. 
The method is applied to 58714 galaxies in CFHTLenS that have spectroscopic redshifts from DEEP2, VVDS and VIPERS.
Using this data we show that the stacked PDF's give an excellent representation of the true $N(z)$ using information from 5, 4 or 3 photometric bands.
We show that the fractional error due to using $N(z_{phot})$ instead of $N(z_{truth})$  is $\leq 1\%$ on the lensing power spectrum ($P_{\kappa}$) in several tomographic bins.  
Further we investigate how well this method performs when few training samples are available and show that in this regime the neural network  slightly overestimates the $N(z)$ at high $z$.
Finally the case where the training sample is not representative of the full data set is investigated. 
%An IPython notebook accompanying this paper is made available here: \url{https://bitbucket.org/christopher_bonnett/nn_notebook}

\end{abstract}


\section{Introduction}
Ongoing and future galaxy  surveys will need to measure the photometric redshift of the order of a billion galaxies (e.g. DES\footnote{http://www.darkenergysurvey.org/}, KiDS\footnote{http://kids.strw.leidenuniv.nl/}, HSC\footnote{http://www.naoj.org/Projects/HSC/}, PAU\footnote{http://www.pausurvey.org/}, J-PAS\footnote{http://j-pas.org/}, LSST\footnote{http://www.lsst.org/}, Euclid\footnote{http://sci.esa.int/euclid/}, WFIRST \footnote{http://wfirst.gsfc.nasa.gov/}).
The estimation of the correct redshift distribution is of great importance for several cosmological probes.  
Baryonic acoustic oscillations and weak gravitational lensing are two probes that have strong potential to constrain dark-energy \citep{DETF} for which accurate redshifts are essential.
For example, in weak lensing the strength of the lensing signal
is directly dependent on the distance between the observer, the lenses and the sources. 

In the literature there are two main methods of measuring photometric redshift, the first being template fitting methods the other being empirical training methods (see\cite{phat} for a comparison of the different methods).
Template fitting methods rely on fitting empirical or synthetic galaxy spectra convolved  with observed filters and telescope response to the observed magnitudes in the survey
\citep[e.g.][]{BPZ,LEPHARE,ZEBRA}.
Empirical methods use a set of spectroscopic  training redshift to calibrate an algorithm to learn a mapping between the observed magnitudes and the redshifts of the galaxies. 
There are a slew of machine learning algorithms that have been applied to the  photometric redshift problem: neural networks \citep[e.g.][]{ANNZ}, boosted decision trees \citep[e.g.][]{AZ}, random forests \citep[e.g][]{carrasco}, gaussian processes \citep[e.g][]{way2009}, self-organised maps \citep[e.g.][]{geach2012}, spectral connectivity analysis \citep[e.g.][]{freeman2009}, support vector machines\citep[e.g.][]{svm2005} and quasi newton algorithm \citep[e.g.][]{c2012}.
From the machine learning methods TPZ \citep{carrasco}, ArborZ \citep{AZ} and  the method described in \cite{wolf} are able to provide a PDF for each galaxy. 
This feature is more common in template fitting methods (e.g Le Phare, BPZ, ZEBRA).
Both methods have their advantages and disadvantages. The template fitting methods rely on the assumption that the SED templates are a true representation of the observed SED's, which is not necessarily the case.  For training based methods it is essential that the training data is a true representation of the full survey data.  When the full galaxy sample contains galaxies that are not or sparsely represented in training data then the results for those subsamples may be unreliable.
Several  methods have been proposed in the literature to infer the $N(z)$ by cross-correlating with a spectroscopic reference population.
Most methods rely on cross-correlation at large scales \citep[e.g][]{new2008,ben2010,white2013} while \cite{menard} advocate adding small scales information.
\cite{jj} propose increasing the redshift precision by imposing an isotropy and two-point correlation prior in a Bayesian analysis after the initial redshift have been measured using methods mentioned above.  

All methods need a spectroscopic set to train, calibrate and/or validate and thus will also depend on any systematics present  in the spectroscopic set, this has been studied in \citet{cunhab, cunhaa}.
An overview and discussion of the spectroscopic requirements for future surveys is given in \cite{new2013}. 

The paper is organised as follows: in  Sect. \ref{rr} we give a brief overview of neural networks and explain the methodology
of obtaining a PDF for each galaxy using a NN.
Then we show how well the method works on CFHTLenS data in Sect. \ref{gen}.
Further we investigate the performance when few training samples are available (Sect. \ref{azsx})
Finally we show how the performance is affected  when the NN is trained on a non-representative sample (Sect. \ref{bla}).  
 
\section{Methodology}\label{rr}     
%In machine learning can be divided in two large  categories, unsupervised learning and supervised learning.  
The photometric redshift problem is categorised as a supervised learning problem in machine learning as opposed to unsupervised learning. 
We have a set of labeled training data with some properties and we wish to infer the labels on data  for which we only have the properties. 
The properties are usually called features, in the case of photometric redshift this usually consists of photometric information but also can contain information like galaxy size.  
A further division is made into regression and classification problems. 
In classification problems the labels are discrete values while the labels in regression problems take on continuos values.      
In the light of the photometric redshift problem neural networks have been used to obtain the most likely photometric redshift as a regression problem \citep[see][]{ANNZ}.
In this work we show how one can use a classification NN to estimate the redshift distribution of a galaxy sample. 

\subsection{Neural Networks}
The simplest form of a NN consists of several ordered layers of perceptrons, which is  called a multi-layer percepetron (MLP). A perceptron is an algorithm that maps an input vector $\mathbf{x}$  to a scalar. We follow the notation from \cite{MacKay}. 
\begin{equation}
f(\mathbf{x};\mathbf{w},\theta) = \theta + \sum_i^n w_i x_i
\end{equation}     
Here $\mathbf{w}$ = ${w_i}$ are the weights of the network and $\theta$ is known as the bias.
A three layer perceptron has one input layer one hidden layer and  one output layer.
Given an input layer with $l$ nodes, a hidden layer with $j$ nodes and final layer with $i$ output nodes then the outputs of the hidden layer are given by
\begin{equation}
h_j = g^{(1)}(f_j^{(1)})  \text{ with }  f_j^{(1)} = \theta_j^{(1)} + \sum_l w_{jl}^{(1)}x_l.
\end{equation} 
The output layer produces  
\begin{equation}
y_i = g^{(2)}(f_i^{(2)}) \text{ with } f_i^{(2)} = \theta_i^{(2)} + \sum w_{ij}^{(2)}h_j. 
\end{equation}
Here $ g^{(1)}(x) = 1/(1+e^{-x}) = sig(x)$, this is known as the sigmoid function and
$g^{(2)}(x) = x$.  
A more complex architecture can be achieved by stacking several hidden layers. A network with more hidden layers can model higher complexities but as a down side they tend to over-fit if one is not careful to stop the training at the appropriate time. 
\begin{figure}
\centering
\includegraphics[scale=0.39]{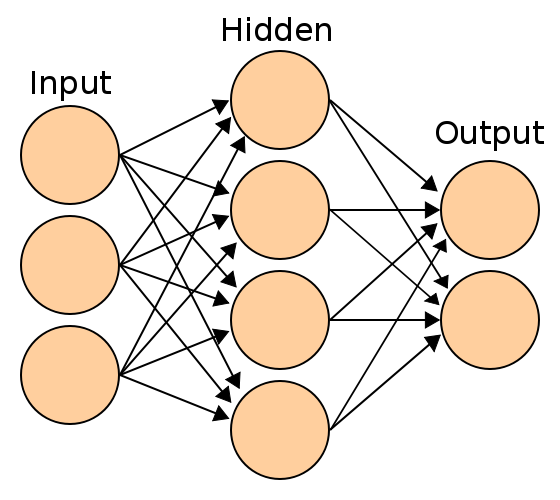}
\caption{The representation of a 3 layer neural neural net with 3 input nodes, 4 hidden nodes and 2 output nodes. Image courtesy of Wikimedia Commons.}
\label{NN_fig}
\end{figure}
Figure \ref{NN_fig} is a diagram of a 3 layer network with 3 inputs, 1 hidden layer with 4 nodes and  an output layer with 2  nodes.

\subsection{Regression and classification networks}
The NN used in this work (see Sect. \ref{soft}) considers the network parameters to be random variables with the following posterior  
\begin{equation}
 \mathcal{P}(\mathbf{a}; \alpha,\mathbf{\sigma}) =  \mathcal{L}(\mathbf{a}; \mathbf{\sigma}) \times  \mathcal{S}(\mathbf{a};\alpha)
\end{equation}
Here $\mathbf{a}$ denotes all the network parameters (i.e the weights and biases).
$ \mathcal{L}$ is the likelihood that depends on $\mathbf{a}$ and the hyper parameter
$\mathbf{\sigma} =\{ \sigma_i \}$, this is the standard deviation of the outputs. 
$ \log  \mathcal{S}(\mathbf{a},\alpha) = \frac{\alpha}{2} \sum_i a_i^2 $ is the prior with $\alpha $ acting as a regularisation parameter and is usually referred to as the weight decay rate.
The prior favors small values of $\mathbf{w}$ and decreases the tendency of the network to overfit on the training data.
The likelihood for a regression network differs from  the one of  a classification network.
For a regression network the log-likelihood is given by 
\begin{equation}
\begin{split}
\log  \mathcal{ L } (\mathbf{a};\mathbf{\sigma}) = - \frac{K \log(2 \pi)}{2} - \sum_{i=1}^{N} \log (\sigma_i) \\
 - \frac{1}{2}\sum_{k=1}^K \sum_{i=1}^N \left[ \frac{t_i^{(k)} - y_i(\mathbf{x}^{(k)};\mathbf{a})}{\sigma_i} \right]^2
\end{split}
\end{equation}
Here N is the number of outputs, K is the amount of training samples.
$y_{i}$ $(\mathbf{x}^{(k)};\mathbf{a})$  is the prediction of $ t^{(k)} $ using $ x^{(k)} $ as inputs.
When a classification problem contains mutually exclusive classes only, a classification network can give the probability that an object belongs to a certain class. 
This is done by applying a \textit{softmax} transformation to the output layer. 
\begin{equation}
y_i(\mathbf{x}^{(k)};\mathbf{a}) \rightarrow \frac{\exp[y_i(\mathbf{x}^{(k)};\mathbf{a})]}{\sum_{j=1}^N \exp[ y_i (\mathbf{x}^{(k)};\mathbf{a})]}
\end{equation}
This transforms the output to all positive values that add up to 1 and as such can be interpreted as a PDF.
The likelihood in the case of classification  NN is given by the \textit{cross-entropy} of the targets  and the \textit{softmax} transformed outputs .
The cross entropy of two PDF's is at its minimum when the two functions are equal.
\begin{equation}
\log  \mathcal{ L } (\mathbf{a};\mathbf{\alpha}) = - \sum_{k=1}^K \sum_{i=1}^N t_i^{(k)} \log y_i(\mathbf{x}^{(k)};\mathbf{a})
\end{equation}
The NN uses a 2nd-order optimisation method based on the conjugate gradient algorithm, we refer the reader to \cite{graff} for a detailed description of the network training algorithm.

\subsection{Convergence}

If the network architecture is complex enough eventually it will start to over-fit the data. 
To combat this behaviour the data  is divided into 2 sets, a training set and a validation set. 
The NN is fed  the training set and the squared error is calculated  on the training set and the validation set after each training iteration.
When the squared error of the validation set starts to rise while the error on the training set is still descending then this indicates that the network is starting to over-fit. 
At this point the algorithm stops optimising and the best network parameters are returned to the user.

\subsection{Acquiring the probability density function for a galaxy using a classification NN}
Instead of using a regression network to get the best possible photometric redshift we use a classification network to estimate the probability that a galaxy is in a certain redshift bin. 
Given that a galaxy cannot be in more than one redshift bin at the same time a classification NN with a \textit{softmax} transformation is ideally suited for this purpose. 
Before training the network we bin our data in $n$ redshift bins, these bins are our classes. 
When training the network, the features, in this case being the magnitudes and the magnitude errors, and the classes  are fed to the NN.
The class consists of one number between 0 and $n-1$.  
The NN outputs $n$ values between $[0,1]$ one for each class that sum up to $1$, these can be interpreted as the probability that the galaxy resides in that class which in this case is a redshift bin.

\subsection{Software}\label{soft}
Two public NN software packages where used in the initial stages of  this work: $SkyNet$ \footnote{http://ccpforge.cse.rl.ac.uk/gf/project/skynet/} \citep{graff} and the $PyBrain$ \footnote{http://pybrain.org/} \citep{pybrain} implementation of neural networks, both gave consistent results. All the figures and data shown in this paper where obtained with the MPI version of $SkyNet$.
%An IPython notebook \citep{ipython} calling $SkyNet$ accompanying this paper is made available here: \url{https://bitbucket.org/christopher_bonnett/nn_notebook}.

\section{Application to the CFHTL\lowercase{en}S dataset}\label{gen}
We apply our method to the CFHTLenS \footnote{http://www.cfhtlens.org/} data set \citep{erben,heymans} to see how well the method performs in estimating the $N(z)$. 
The photometric information used is the one as measured in \cite{HH} (H12 from here on). 
All magnitudes are in CFHTLenS are $AB$ magnitudes.  
A comparison with the photometric redshifts from H12 can be found in appendix \ref{app_bpz}.
CFHTLenS is based on the Wide component of the Canada-France-Hawaii Telescope Legacy Survey (CFHTLS).
CFHTLS was observed in 5 broad bands $u^{*},g',i',r'$ and $z'$ in 4 independent fields (W1, W2, W3 and W4) with the following mean observing times per band:
\begin{itemize}
\item $u^{*}$ band  $3000$ $s$
\item $g'$ band  $2500$ $s$
\item $r'$ band  $2000$ $s$
\item $i'$ band  $4300$ $s$
\item $z'$ band  $3600$ $s$.
\end{itemize}

\begin{figure}
\centering
\includegraphics[scale=0.45]{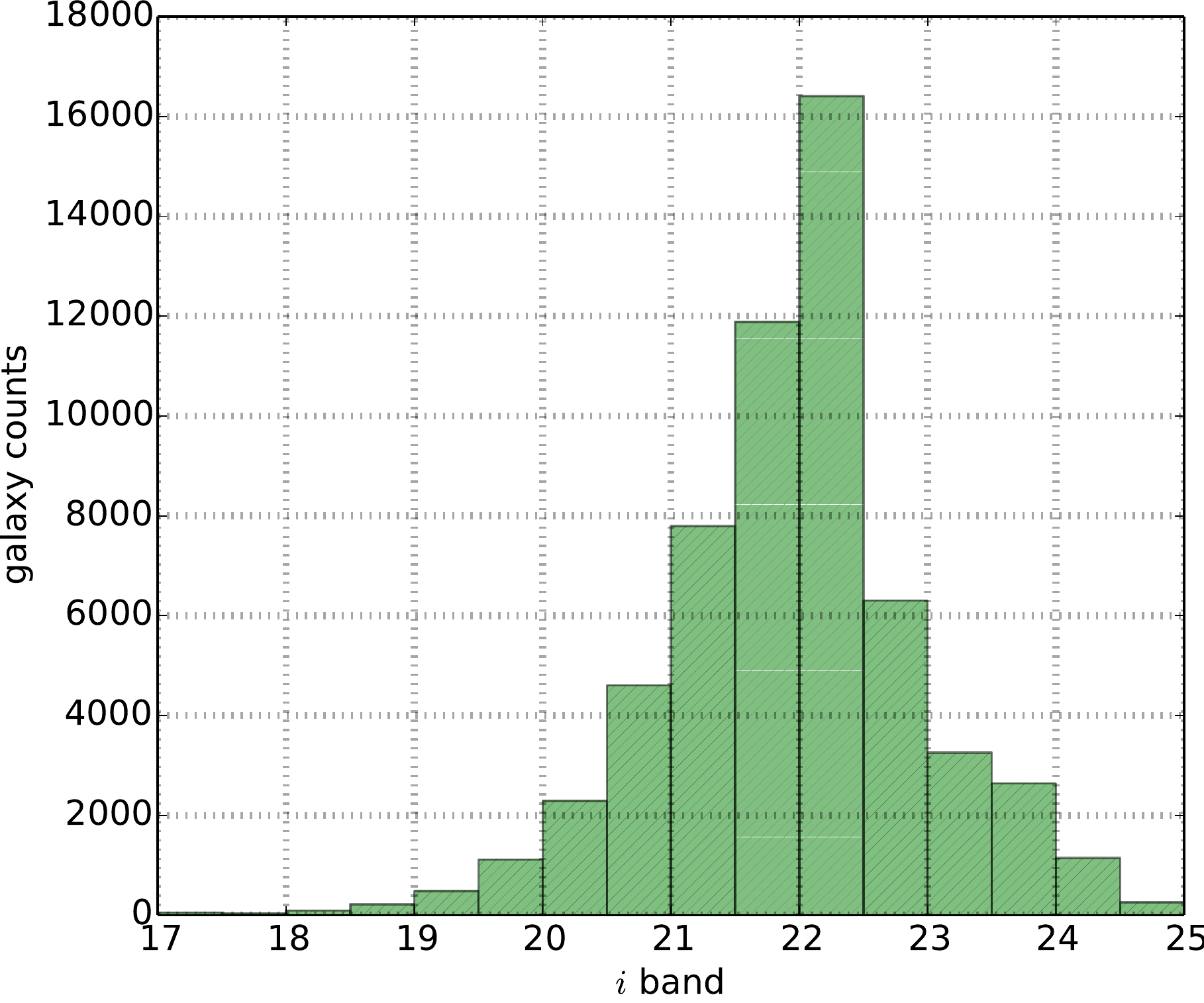}
\caption{The $i'$ band magnitude distribution in the CFHLTenS catalogue with spectroscopic redshift.}
\label{plot_i}
\end{figure}

We use the spectroscopic redshifts  from the following surveys : VVDS deep \citep{vvds}, VVDS F-22 \citep{vvdsf22}, DEEP2 \citep{deep2} and VIPERS \citep{vipers} that lie within the CFHTLS fields.  
VVDS deep is located in the W1 field  and was selected in the magnitude range $17.5  \leq I_{AB} \leq 24.0 $.
The DEEP2 survey is located in the W3 field where objects where selected in $18.5     \leq R_{AB} \leq 24.1$ and selected to maximise the amount of galaxies over stars.
The VVDS-F22 survey lies within the W4 field and galaxies are selected in  $17.5  \leq I_{AB} \leq 22.5 $.
VIPERS survey lies within the W1 and W4 fields, targets where colour selected to maximise the number galaxies in  $ 0.5 < z  < 1.2 $ down to $i_{AB} <22.5$.
In Figure \ref{plot_i} we show the  CFHTLenS $i'$ band magnitude  distribution for all the 58714 matched galaxies  with  $0 < z \leq  2.0$. 
The $z=2.0$ limit guarantees that any galaxy with a higher redshift will be misclassified, we propose a possible solution to this problem in Sect.\ref{dd}.   
We perform the analysis using 40 redshift bins within the redshift range.
 This provides redshift information at $\Delta z =0.05$ resolution.
It is advised to have equal amount of nodes in the last hidden layer as there are classes to classify, in this case this is 40.
Therefore the NN is trained using 3 hidden layers with respectively  20, 40 and 40 nodes per layer. 
No optimisation of the network architecture or parameters is attempted in this study. 
The data set is subdivided into two sets, the training set and the validation set.  The division is 70\% for training and 30\% for the validation set. 
This leads to 41100 training galaxies and 17614 validation galaxies.

\subsection{$N(z)$ estimates using 5, 4 or 3 photometric bands}\label{qwert}

In this subsection we investigate how well the stacked PDF's perform in approximating the true $N(z)$ in  8 tomographic bins.  
The division of the tomographic bins can be seen in Table \ref{std_table}.
Tomographic bin 2 to 7 correspond to the same tomographic  bins that \cite{heymans2012} used for the tomographic lensing analysis in CFHTLenS.
When selecting the galaxies in photometric redshift space there are several options on which $z_{phot}$ to select. 
Some options are to take the mode or mean  of the PDF.
Another option would be to run the NN in regression mode and use the output as the $z_{phot}$ estimate.  
Using a regression NN, the mean and the mode of the PDF as $z_{phot}$ showed similar results in the initial stages of this work. 
In the rest of this work we use the mode of the PDF as $z_{phot}$ to select galaxies and leave a full analysis of which $z_{phot}$ is best to use for future work.
We perform the analysis using all five bands ($u^{*},g',i',r'$ and $z'$) and then consecutively drop the $u^{*}$ and $i'$ band.
The $u^{*}$ band is dropped as several wide field surveys do not observe in this band. 
The $i$ band is dropped for two reasons, first it is the deepest band and such removing it tests the method in an extreme case.  
Secondly to generally  investigate how well the method performs when only 3 bands are available.
The 4 and 3 band virtual surveys respectively use $80\%$ and $52 \%$ of the 5 band observing time.   
A similar exercise could be done with any other permutation of 4 and 3 bands.

To get an estimate of the errors we make 50 random training and validation sets.
We thus train 50 networks and obtain 50 PDF's for each galaxy allowing us to use the mean as best PDF$_{phot}$ and use the variance between the sets to estimate the errors (i.e cross-validation).
We note that this error estimate is the measurement error given the galaxy features and the variation of the features in the spectroscopic set it does not take into account spatial variance usually referred to as cosmic variance. 
\begin{figure*}
\centering
\includegraphics[scale=0.540]{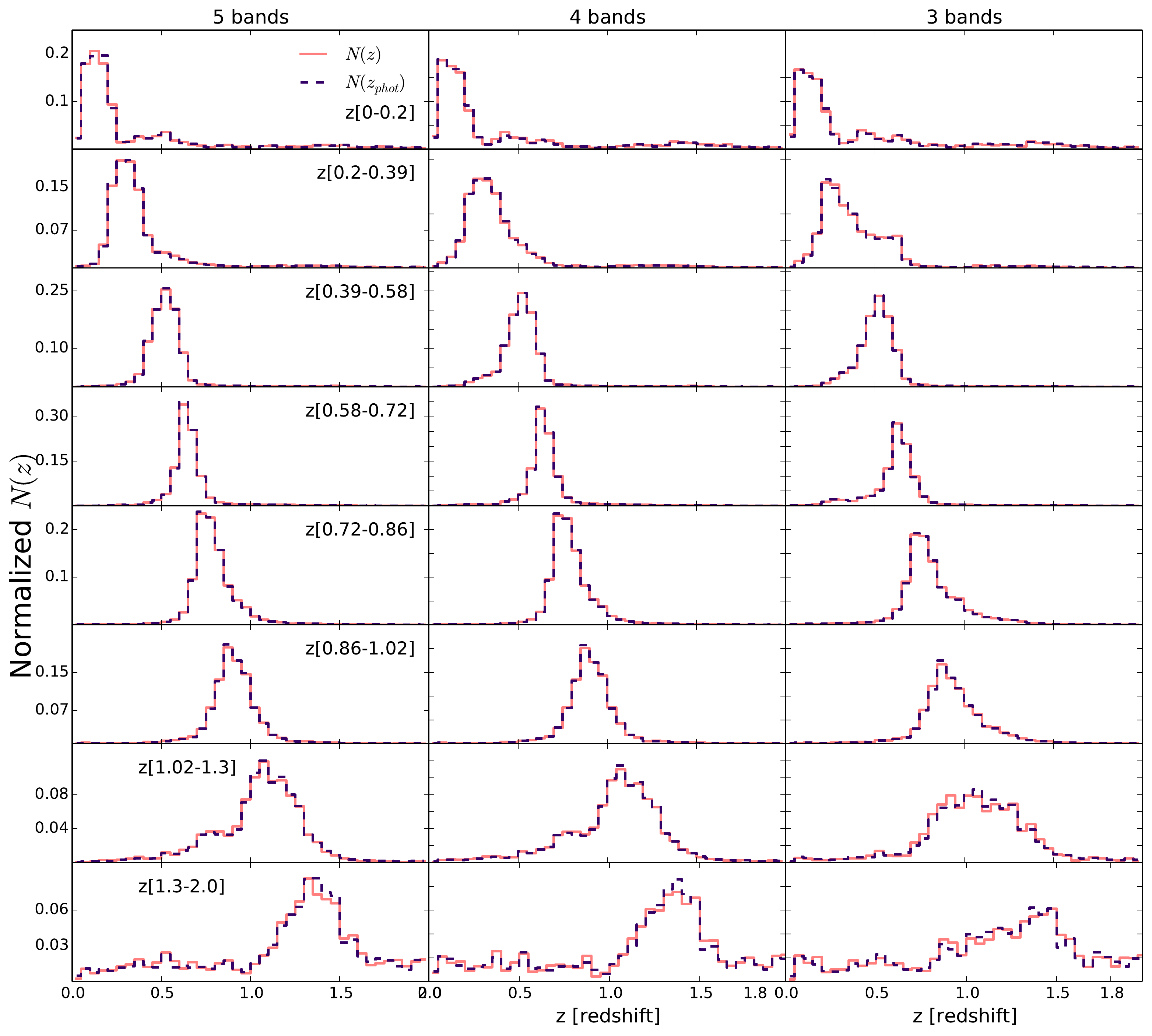}
\caption{The figure shows the mean of the 50  validation samples for the NN $N(z_{phot})$  (blue-dashed line) and the true $N(z)$ (solid-red line) of the validation set. 
The left panels show the results  using $u^{*},g',r',i',z'$ photometric information. The middle panels show the same for $g',r',i',z'$ with the right  panels showing the results using just $g',r',z'$ photometric information. No error bars are shown due to most of them not being visible in the plot.}  
\label{plot_5}
\end{figure*}
In Figure \ref{plot_5} we plot the mean $N(z_{phot})$ and $N(z)$ of 50 validation sets as learned by the NN in the case of using 5, 4 or 3 photometric band information as features.
The true $N(z)$ is estimated by binning the spectroscopic redshifts for the selected galaxies in bins of the same width as the NN output ($\Delta z =0.05$). 
\begin{figure*}
\includegraphics[scale=0.52]{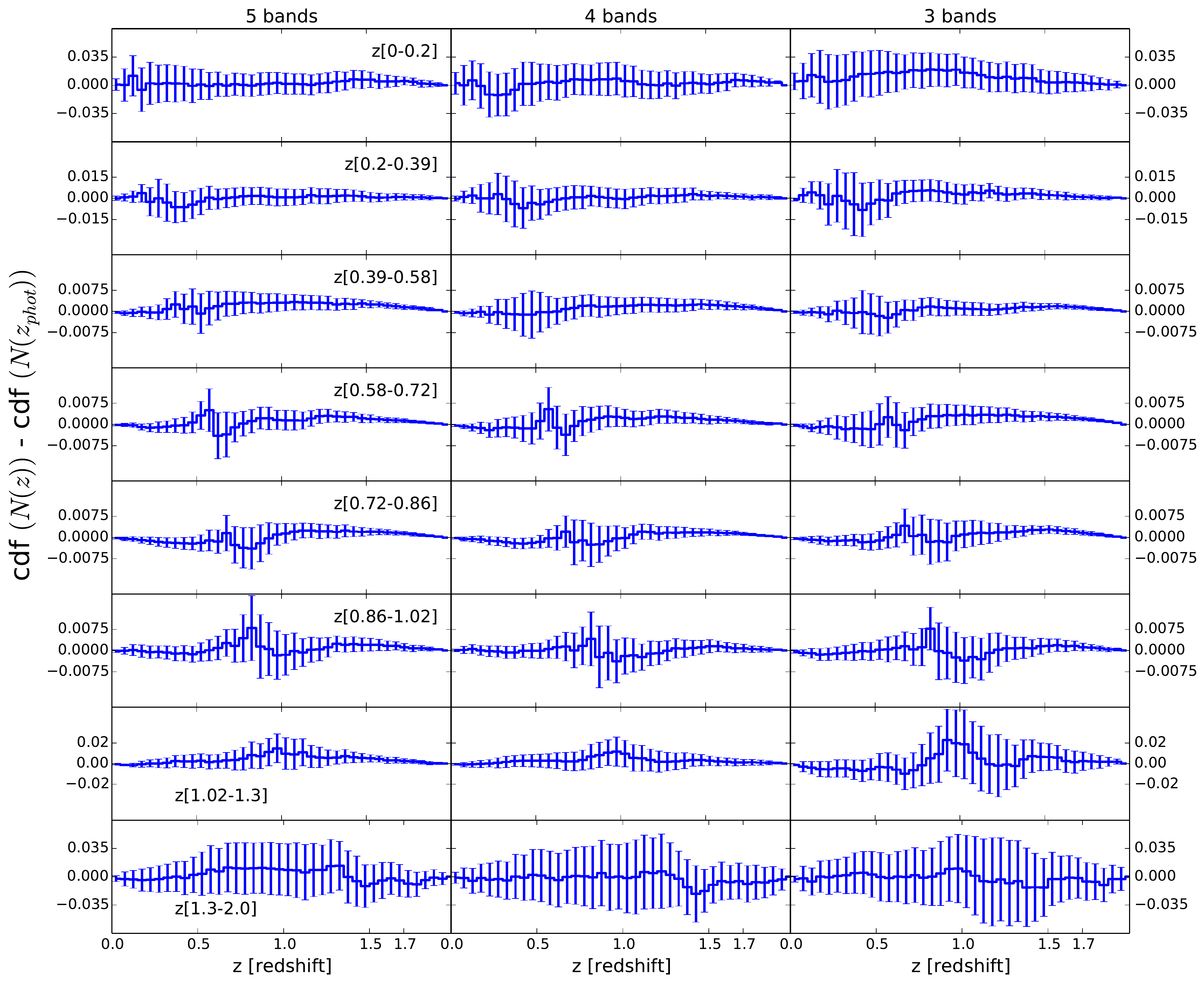}
\caption{The left panels show the difference between the cumulative $N(z_{phot})$ and the   cumulative $N(z)$ using $u^{*},g',r',i'$ and $z'$ photometric information. The error bars are standard deviation between the 50 validation samples. The middle panels show the same for $g',r',i'$ and $z'$ and the right most panels show the results using just $g',r',z'$ information.
For low and high redshift bins the errors increase significantly from the 5 band case to the 3 band case. For intermediate redshift the error increase is minimal. Note that the y-axis is not the same for all redshift bins.} 
\label{plothuh}
\end{figure*}
In most of the panels the difference between $N(z)$ and $N(z_{phot})$ cannot be perceived.
We interpret this as the NN learning the degeneracies for the galaxies at different redshifts given the input features. 
The loss of information from dropping photometric information is most notable in the resulting width of the galaxy selection. 
The width of the redshift distribution increases with less bands, this is most notable in the low and high redshift bins. 
This  can also be seen in the transition matrices that are shown in Appendix \ref{app_trans} for the 3 cases.
A more quantitative way to judge the performance of the NN is plotted in Figure \ref{plothuh}, it shows the difference between the cumulative $N(z)$ and the cumulative $N(z_{phot})$.
The errors are given by the standard deviation between the 50 validation samples.   
Figure \ref{plothuh} shows that the loss of photometric bands leads to larger errors in the estimation of the cumulative distribution for most redshift bins.  
In Table \ref{std_table} we show the maximum size of the standard deviation for all the 8 tomographic bins as a function of the photometric bands used. 
The largest increase in maximum error happens in the first and last two tomographic bins.
In the other bins there is either a minimal increase or the values are very similar. 
In tomographic bin 6 the maximum error decreases when using less photometric bands. 
This could be within the noise (i.e. within the error on the error bar) or the fact that for this tomographic bin the NN does learn the degeneracies better when less photometric band information is fed to the NN. 
The more features fed to a NN the more training data it needs to learn a correct mapping from the features to the correct labels.
At high $z$ in tomographic bin 3 through 5 in Figure \ref{plothuh} the difference between the cumulative functions is not consistent with zero. The values are of the order $\sim 0.0003 \pm 0.0002$. 	
The NN tends to put slightly more probability at high $z$ than the truth, this can also be seen in the transition matrices in Appendix \ref{app_trans} (\ref{tm5}, \ref{tm4} and \ref{tm3}) where, for example, when selecting galaxies in 
tomographic bin 3 a fraction of 1.4\% $\pm$ 0.2\% are in tomographic bin 8 while the NN predicts 1.8\% $\pm$ 0.2\% of the galaxies to be there when using 5 band photometric information. 
 \begin{table}
\center
\hspace{30mm}The maximum standard deviation 
\begin{tabular}{l*{3}{c}r}
\hline
$z$ bin & $z -selection$ & 5 bands & 4 bands & 3 bands  \\
\hline
bin 1 &0.0  $ < z$ $\leq$  0.2& 0.0270  & 0.0280  & 0.0336 \\
bin 2 &0.2  $ < z$  $\leq$  0.39& 0.0134  & 0.0161 & 0.0223  \\
bin 3 &0.39 $ < z$ $\leq$   0.58& 0.0069  & 0.0083 & 0.0076  \\
bin 4 &0.58 $ < z$  $\leq$  0.72 & 0.0081  & 0.0077 & 0.0079  \\
bin 5 &0.72$  < z$  $\leq$  0.86 & 0.0073 & 0.0079  & 0.0079  \\
bin 6 &0.86 $ <  z$ $\leq$  1.02 &  0.0115 & 0.0109  &  0.0100 \\
bin 7 &1.02$  < z$ $\leq$  1.3  &  0.0153 & 0.0141  & 0.0334 \\
bin 8 &1.3  $  < z$  $\leq$  2.0 & 0.0343  & 0.0459  &  0.0545\\ 
\end{tabular}
\caption{The maximum standard deviation value of the difference of the cumulative $N(z)$ and $N(z_{phot})$ estimated on the 50 validation samples  for each of the redshift bin given the amount of photometric bands used.}
\label{std_table}
\end{table}

\begin{figure*}
\includegraphics[scale=0.53]{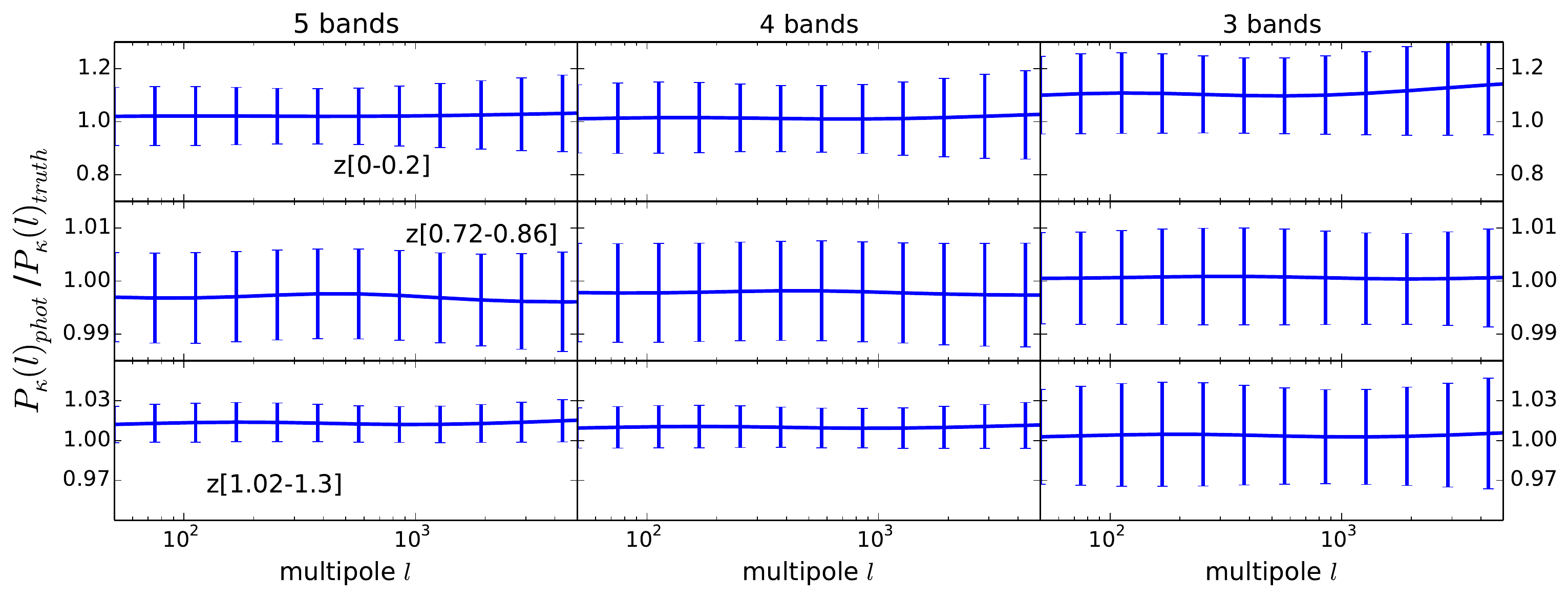}
\caption{The left panels show the ratio between $P_{\kappa}(l)_{phot}$ and $P_{\kappa}(l)_{truth}$ using $u^{*},g',r',i'$ and $z'$ photometric information for tomographic bin 1, 5 and 6 for $ 50 \leq l \leq 5000$. 
The error bars are standard deviation between the 50 validation samples. 
The middle panels show the same for $g',r',i'$ and $z'$ and the right most panels show the results using just $g',r',z'$ information. 
Note that the y-axis is not the same for all redshift bins.} 
\label{PK}
\end{figure*}

\subsection{$P_{\kappa}(l)$ Estimation}\label{PK1}
In this section we see how the theoretical lensing power spectrum $P_{\kappa}(l)$ from $N(z_{phot})$ compares to the $P_{\kappa}(l)$ of the true $N(z)$ (see \cite{bs} for a review for weak gravitational lensing theory).
We calculate the lensing power spectra for the 50 validation  $N(z)$'s and $N(z_{phot})$'s in all 8 tomographic bins that where obtained using 5, 4 or 3 band photometric information. 
This was done with the publicly available $Nicea$ code \footnote{http://www2.iap.fr/users/kilbinge/nicaea/} \citep{nc}.
We assume  a $\Lambda$CDM cosmology using the \cite{sm2003} fitting function
($\Omega_m = 0.25$, $\Omega_b = 0.044$, $h=0.7$, $n_s=0.95$, $\Omega_{\Lambda}0.75$, $\sigma_8 = 0.8$ and $w=-1$). 
To asses the performance we take the ratios of $P_{\kappa}(l)_{phot}$ and  $P_{\kappa}(l)_{truth}$ with range of  multipole $50 \leq l \leq 5000$.  
The errors are given by the standard deviation of the 50 samples.
Figure \ref{PK} shows the results for the ratios for selected tomographic bins as a function of $l$. Table \ref{tpk} summarises the information for all the bins averaged over all $l$ values.
All the ratios are consistent with 1.
In tomographic bin 3, 4, 5 and 6 we are able to obtain an  fractional error on $P_{\kappa} \hspace{1mm} \pm  \leq1\%$  when using 5 band photometry information. 
When using 4 bands this is true  for tomographic  bin  5 and 6.   
When using 3 bands this is true only for tomographic  bin 5.   
In Appendix C we provide a fitting function to the fractional error on $P_{\kappa}$ as function on galaxy density in the tomographic bin.   
\begin{table}
\center
\hspace{20mm} Ratios of $P_{\kappa}(l)_{phot}$ /$P_{\kappa}(l)_{truth}$.
\begin{tabular}{l*{3}{c}r}
\hline
$z$ bin    & 5 bands & 4 bands & 3 bands  \\
\hline
bin 1 &   1.02   $\pm$ 0.11     &  1.02   $\pm$ 0.14   &  1.11    $ \pm $  0.16\\
bin 2 &   0.996 $\pm$ 0.047  &  0.996 $\pm$ 0.036 & 1.007   $ \pm $  0.042\\
bin 3  &  1.004 $\pm$ 0.009  &  1.001 $\pm$ 0.012 & 0.998   $ \pm $  0.013\\
bin 4  &   1.001 $\pm$ 0.010 &  1.000 $\pm$ 0.012 & 1.001   $ \pm $ 0.012\\
bin 5  &   0.996 $\pm$ 0.009 &  0.998 $\pm$ 0.009 & 1.001   $ \pm $ 0.009 \\
bin 6  &   1.000 $\pm$ 0.009 &  1.000 $\pm$ 0.009 & 0.999   $ \pm $ 0.015 \\
bin 7  &   1.013 $\pm$ 0.014 &  1.010 $\pm$ 0.016 & 1.003   $ \pm $ 0.037 \\
bin 8  &   1.010 $\pm$ 0.056 &  1.002 $\pm$ 0.079 & 1.00     $ \pm $ 0.08  \\ 
\end{tabular}
\caption{The ratios of $P_{\kappa}(l)_{phot}$ over  $P_{\kappa}(l)_{truth}$ averaged over the entire $l$ range ($50 \leq l  \leq5000$) for the 8 tomographic bins using 5, 4 or  3 band information.
The error is mean error bar size over the  full $l$ range. 
All ratios are consistent with 1.
See Figure \ref{PK} for a $l$ dependent plot}
\label{tpk}
\end{table}

\section{Performance as a function of  number of training samples}\label{azsx}
In this section we vary the number of training samples and see how the method performs when a few but fair (i.e random subsample) training samples are available. 
This test is performed using all 5 photometric bands.
We fix the validation set to a random sub-sample of 20000 galaxies this leaves us  with a parent training set of 39388 galaxies.  
From this parent training catalogue we make various training sets. 
We vary the training sample size between 250  to 3500 galaxies and choose the galaxies  so that no galaxy appears twice in a training set for a certain training sample size, i.e. sampling without replacement. 
See Table \ref{t_table} for details on the training sets.
\begin{table}
\center
\begin{tabular}{l*{7}{c}r}
\hspace{25mm}Ratio of $N(z_{phot})/N(z)$ \\
\end{tabular}
\begin{tabular}{l*{4}{c}r}
\hline
Training set size & \# sets  & z=[1.6-1.65]  &  $z=[1.95-2.0]$\\
\hline 
250 & 157   &   3.5  $\pm$ 1.4  &   8.3 $\pm$ 2.6\\
500 & 78     &     2.6 $\pm$ 0.9 & 6.5 $\pm$ 2.3 \\
 1000 & 39  &  1.9  $\pm$  0.8  & 4.6  $\pm $1.9 \\
1500 & 26   &    1.7 $\pm$ 0.8  &   3.7  $\pm$ 1.3\\
 2000 & 19  &  1.4  $\pm$ 0.7   &  3.1 $\pm$ 0.8\\
2500 & 15   &   1.5  $\pm$ 0.6    &  3.0  $\pm$ 0.7\\
 3000 & 13  &   1.3 $\pm$ 0.3    &  2.7  $\pm $0.7\\
3500 & 11   &    1.4  $\pm$  0.4  & 2.3 $ \pm$ 0.6 \\ 
%41571 & ### & 1.2 $\pm$ 0.6  &  1.9 $\pm$ 0.9 \\
\end{tabular}
\caption{The first 2 columns depict the training set size and  the number of training sets that fit in the parent training set of  39388 galaxies. 
The last 2 columns are the ratio of  $N(z_{phot})/N(z)$  at high $z$ given the amount of training samples. The error is give by the standard deviation from the results on the different training sets.}
\label{t_table}
\end{table}
\begin{figure*}
\begin{center}
\includegraphics[scale=0.450]{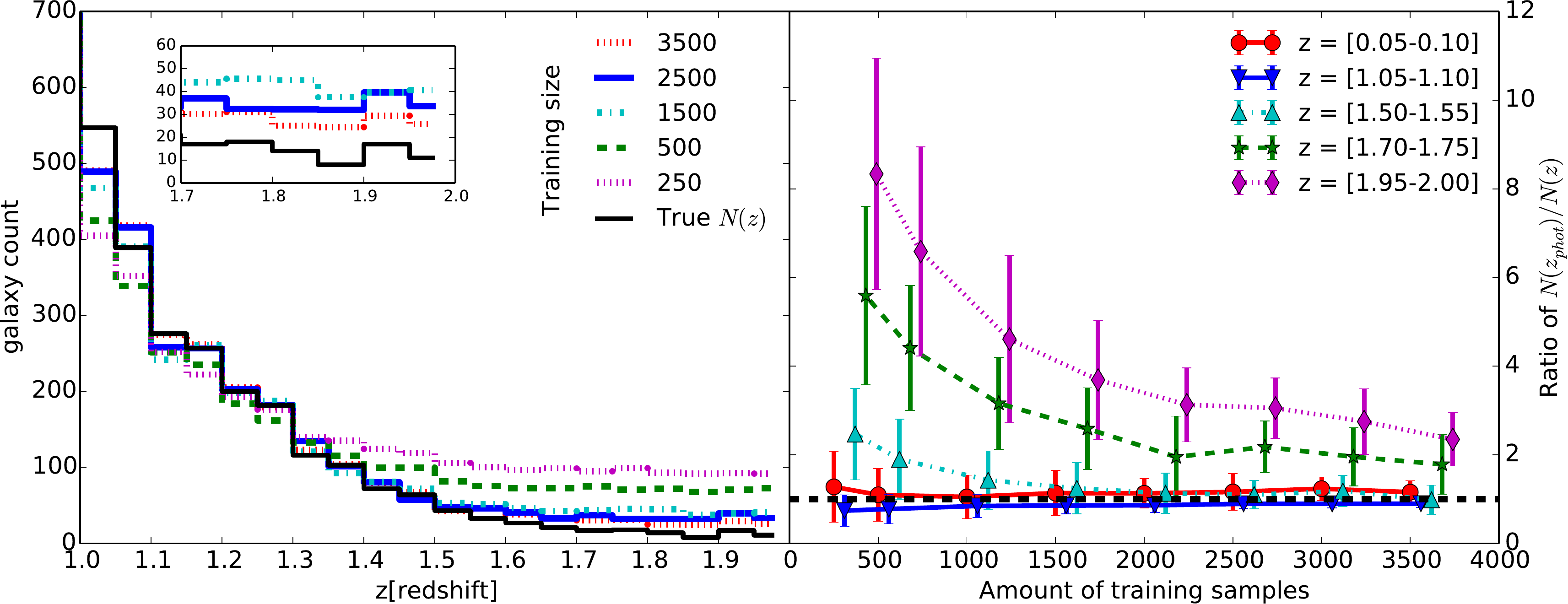}
\caption{The left panel shows the high redshift ($z>1)$ tail of  $N(z)$ and the $N(z_{phot})$ estimates for different training set sizes. Validation set size = 20000 galaxies. Error bars where omitted for sake of clarity. The inset shows the  very high redshift tail where the $N(z_{phot})$ is overestimating the truth. The right panel shows the ratio of $N(z_{phot})$ over $N(z)$ for selected redshift bins with width $\Delta=0.05$ as a function of training size. The black dotted line depicts a ratio $=1$.} 
\label{plot_amount}
\end{center}
\end{figure*}
The left panel of Figure \ref{plot_amount} shows the high redshift tail of the full $N(z)$.
The right panel  shows the estimated values divided by the true values for 5 narrow bins with width $\Delta_z = 0.05$ as a function of  the amount of training samples. 
These redshift bins are the outputs of the NN from the full validation set. 
No galaxy selection took place like in Sect. \ref{qwert}.
The errors are estimated as the standard deviation between the different training sets.
For a small number of training samples $(\sim 250)$  $N(z_{phot})$ is unbiased for $z$ up to $\sim 1$.
When looking at higher redshift  it becomes clear that the $N(z_{phot})$ tends to overestimate the amount of galaxies present. 
The bias diminishes when using more training samples but even for the largest training set number tested (i.e 3500), the estimates at $\sim z >1.6$ are still biased. 
This is most likely due to the small amount of training samples at high redshift. For a training set size of 3500 galaxies  only $\sim$150 galaxies lie beyond $z=1.5$, which is unlikely to be a fair representation of the true galaxy population that resides in the validation set. 
For comparison, when using 70\% of all the spectra as training set there are $\sim 1700$ galaxies with $1.5 < z \leq 2.0$,
the ratio  $N(z_{phot})/N(z) = 1.9 \pm$ 0.8 at $z=[1.95-2.00]$ and consistent with 1 for all other bins.
 
\section{Training on a non representative population}\label{bla}
\begin{figure}
\centering
\includegraphics[scale=0.38]{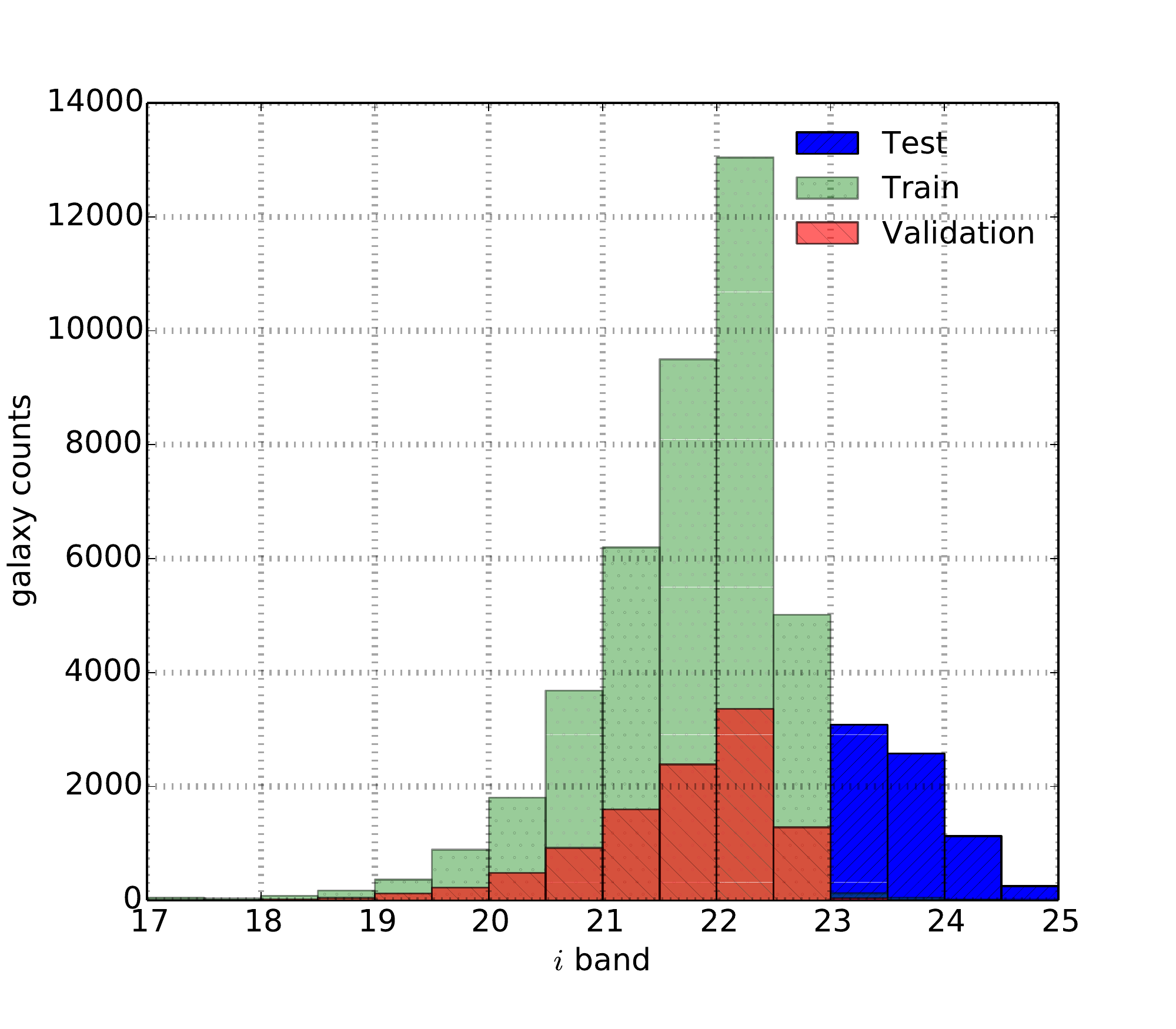}
\caption{The $i'$ band magnitude distribution in the training, validation and test set used in Sect. \ref{bla}}
\label{plotI2}
\end{figure}

In this section we see how well the NN methods performs when we train on a sample that is not representative of the full catalogue using all 5 bands.
To do so we split the data into 3 sets: a training set, a validation set and a test set. 
As before the NN trains on the training sample and uses the validation sample to help decide when to stop training.
Therefore we need that the validation set does not contain galaxy samples that are not present in the training set. 
This would not fairly represent the performance of the NN on a galaxy population for which few to no spectra are available. 
The test set contains galaxies that are not or sparsely represented in the training set. 
As a test set use all the galaxies in the W1 and W3 field with $i > 23.0$.
The validation and training parent set  consists of all galaxies in W1 and W3 that have $i  \leq 23.0$ together with all the galaxies in W4. 
We randomly subdivide that parent set in 70\% training and 30\% validation sets.
There are 207 galaxies with $i >23.0$ in the training set and  7086 in the test set, as a comparison $70\%$ of the full sample with $i > 24.5$  amounts to 215 galaxies.
In Figure \ref{plotI2} the $i'$ band magnitudes are shown for the three sets. Table \ref{w134_table} shows the details of amount of galaxies per set as a function of magnitude.

\begin{figure*}
\centering
\includegraphics[scale=0.39]{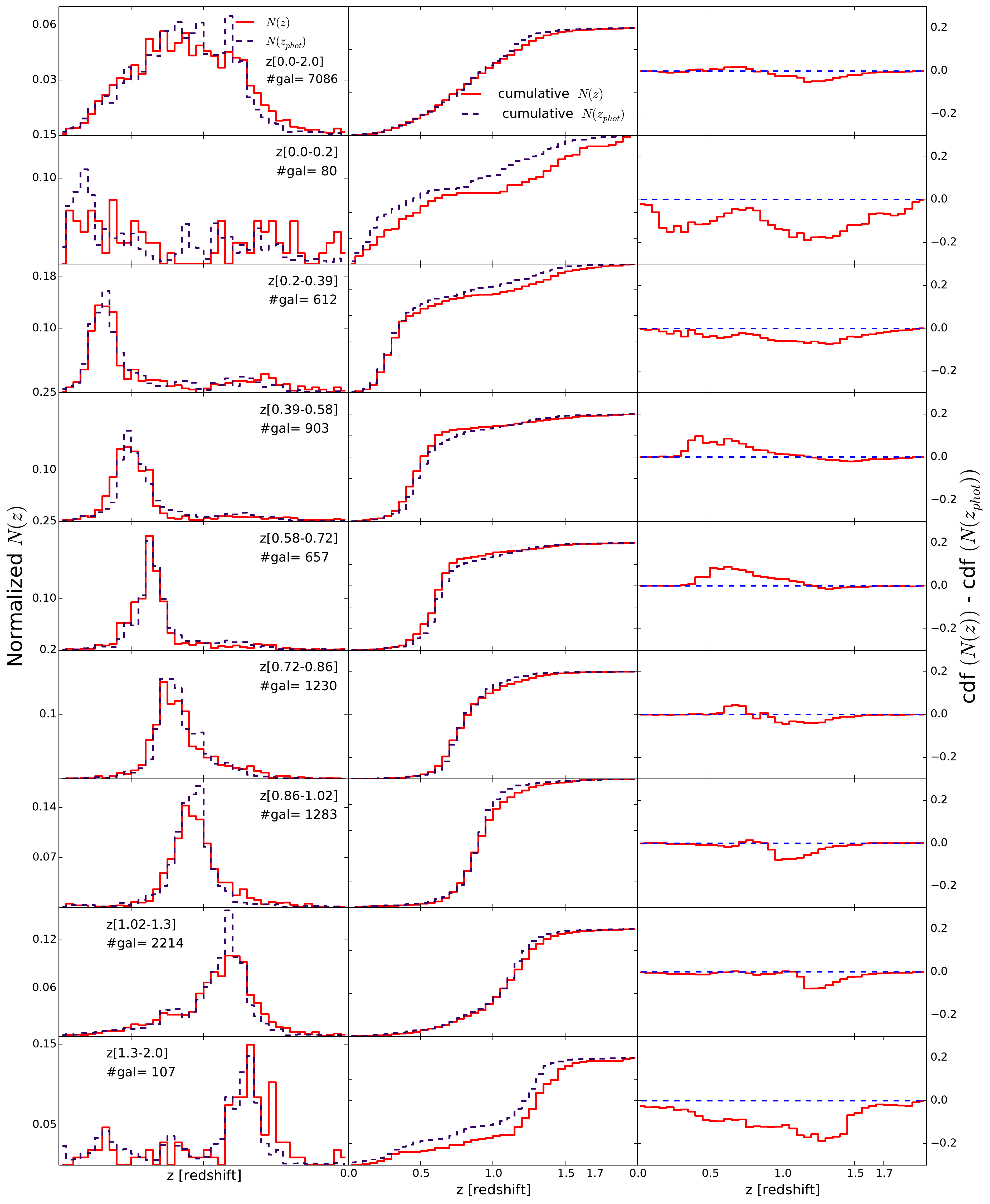}
\vspace{1mm}
\caption{The left panel shows the $N(z)$ and $N(z_{phot})$ for 9 redshift bins of the $7086$ galaxies in the test set. The middle panels shows the cumulative distributions while the third panel shows the difference between the cumulative distributions. The $D$ statistic is equal to the maximum absolute value of the third panel} 
\label{plot_faint2}
\end{figure*}

\begin{table}
\center
\hspace{22mm} Amount of galaxies  \\
\begin{tabular}{l*{3}{c}r}
\hline
$ i $ selection   & training  & validation & test  \\
\hline
$i \leq$ 23.0 &  40892 & 10468  & 0 \\
23.0  $<  i  \leq$ 23.5  &  127 &   41 & 3088 \\
23.5 $<  i \leq $ 24.0   &  48   &   10 & 2579\\
 24.0 $<  i  \leq$ 24.5  &  17   &   3 & 1126\\
24.5 $<  i $                    &  15   &   0 & 295 \\
\end{tabular}
\caption{The amount of galaxies a a function of $i$ band magnitude for the training ,validation and test set.}
\label{w134_table}
\end{table} 
To quantify how well  $N(z_{phot})$ compares to $N(z)$ we use the following metric: the maximum distance between the cumulative distribution function of the true $N(z)$ and $N(z_{phot})$ as done in \cite{cuncha}. 
\begin{equation} \label{ds}
D = \text{max} | \text{cdf}(N(z)) -\text{cdf} (N(z_{phot})) |.
\end{equation}
This test resembles a Kolmogorov-Smirnov(KS) test but is not the same as we perform it on binned data. 
Like the KS test a lower value of D means that $N(z_{phot})$ is a better representation of $N(z)$.
Table \ref{D_table} shows the $D$ values for all redshift bins for the test sample and the $D$ values obtained in Sect. \ref{qwert}.
\begin{table}
\center
\begin{tabular}{l*{3}{c}r}
z-bin   & $D_1$    & $D_2$ & $D_3$ \\
\hline
all   z &  0.051 & 0.0109   &  0.0028  \\
bin 1 &  0.19 &  0.0712  &  0.0115  \\
bin 2 &  0.07 &  0.0324  &  0.0062  \\
bin 3 &  0.09 &  0.0180  &  0.0032 \\
bin 4 &  0.08 &  0.0291  &  0.0050 \\
bin 5 &  0.04 & 0.0260   &  0.0038 \\
bin 6 &  0.07 & 0.0303   &  0.0079 \\
bin 7 &  0.07 & 0.0553   &  0.0149 \\
bin 8 &  0.18 & 0.1053   &  0.0135 \\ 
\end{tabular}
\caption{The $D$ values for different tests. $D_1$ values are those for the test set as studied in Sect \ref{bla}. $D_2$ is the maximum $D$ value of the 50 validation samples when trained on 70\% of the galaxies. $D_2$ is thus the worst result for each redshift bin from Sect. \ref{qwert}.
$D_3$ is the mean $D$ from the 50 validation samples. All 3 cases uses 5 band photometry information.}
\label{D_table}
\end{table}
\begin{figure}
\centering
\includegraphics[scale=0.44]{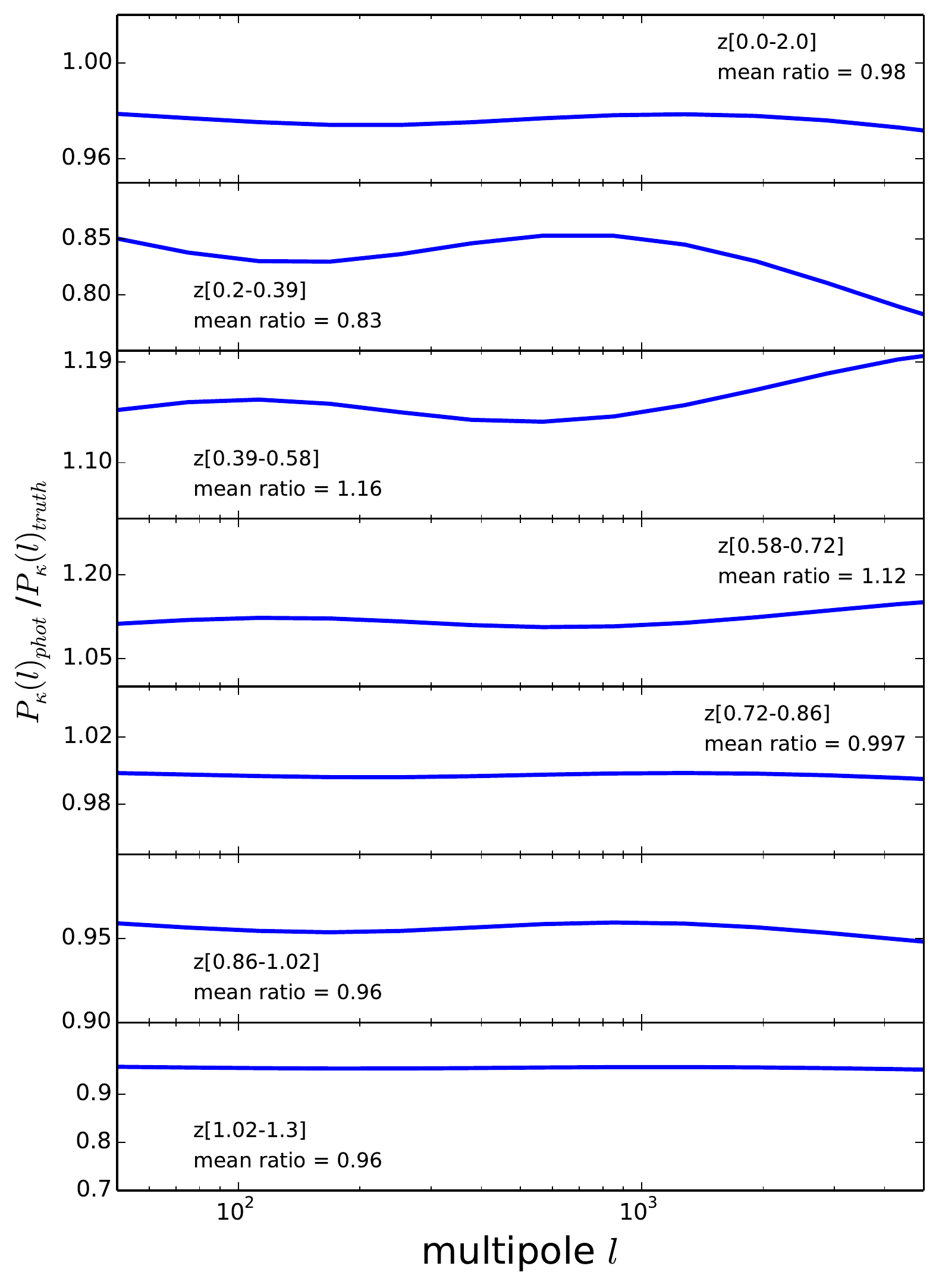}
\vspace{1mm}
\caption{The ratio of $ P_{\kappa}(l)_{phot} / P_{\kappa}(l)_{phot} $ for the test set for $50 \leq l  \leq 5000 $. Tomographic bin 1 and 8 where not plotted as they have ratios of the order $\sim 0.6-0.7$.} 
\label{PK2}
\end{figure}

In Figure \ref{plot_faint2} we show how well the NN is able to retrieve the full redshift distribution compared to the truth.
The NN performs significantly worse than in Sect. \ref{gen}. 
The performance is best in redshift bins where the most training samples are present, showing that the NN has limited but non negligible capacity to extrapolate.
In the lowest and highest tomographic bin the performance is worse although this could be due to the low number statistics in these tomographic  bin. 
These results are therefore as good as one might expect, as the galaxies with spectroscopic redshifts at these faint magnitudes will have brighter magnitude counterparts.  
This is in contrast to the faint galaxies which were targeted by spectroscopic surveys but could not be assigned unique spectroscopic redshifts.
See \cite{new2013} for a discussion on success rate of spectroscopic surveys.
Similar to Sect.\ref{PK1} we show how the lensing power spectrum $P_{\kappa}(l)$ from $N(z_{phot})$  compares to the $P_{k}(l)$ of the true $N(z)$ for the test sample.
Figure \ref{PK2} shows the ratio of the two power spectra. 

\section{Conclusion and Discussion}\label{dd}
In this work we show how to use a classification neural network to estimate the $N(z)$ and apply it to CFHTLenS in  8 redshift bins that are optimised for a tomographic weak lensing analysis. 
By cross-validating on 50 training samples we are able to get an estimate  of the measurement error on $N(z_{phot})$.
We show that  a neural network is able to learn the degeneracies from the data using 5, 4 or 3 photometric band information, $N(z_{phot})$ is excellent approximation of the true $N(z)$.
Leaving out photometric bands leads to wider redshifts selection but has only a small effect on how well the method can recover the $N(z)$.
We show that the NN tends to slightly overestimate  the amount of objects in the high $z$ tail of the distribution by $\sim 0.4 - 0.2 \%$  (see Appendix B for the transition matrixes).
This is likely due to the fact that there are only a few training examples present at high $z$.
This is more clearly seen when we vary the amount of training samples in Sect. \ref{azsx}. As more training samples are added the overestimation becomes smaller.
Adding fake training data from simulated catalogues (\cite{vv}) might be able to mitigate the problem where few training samples are available.
The fractional error on the lensing powers spectrum for 5 bands ranges from $\sim 5\%$ to $ < 1\%$ for all tomographic bins with $z>0.2$.
This range changes to $\sim 8\% $ to $ < 1\%$ when using 4 or 3 band photometric information. 
For the tomographic bin with  $0.0 < z \leq 0.2$ the fractional error is $\sim 11-16 \%$  (see Table \ref{tpk} for the full details). 
The slight overestimation in the high redshift tail is not significant enough to bias the $P_{\kappa}$ estimation. 
Further we show that the NN has some capacity to extrapolate to galaxy samples that our not in the training sample. This appears to work best in redshift ranges where lots of training samples are presents.
More concretely the fractional error on $P_{kappa}$ ranges from $\sim  <1 \% $ to $16\%$ in tomographic bins 2 through 7.

In future research we plan to study the optimal number of galaxies per bin (i.e. per class) as such to optimise the redshift resolution without biasing the $N(z)$ estimation. 
In regions where many training samples are present more fine grained redshift information can be obtained while for regions where training samples are scarce one could use wider bins.
Taking this to the extreme the NN could be trained with one or several very wide redshift bins at very high $z$  that contains all the training galaxies above some $z$ threshold. 
This could alleviate the case where high $z$ galaxies get misclassified as is guaranteed to happen in the current setup. 
Further  it is unclear how much information is carried in the size and ellipticity of a galaxy for this sample using this method.
\cite{ANNZ} showed that adding size and concentration index information to  the regression network ANNZ  showed a 3\% improvement on $\sigma_{rms}$.
For example, one would expect that for edge on galaxies that are redder due to dust, the ellipticity of the galaxy provides useful information that NN can use to break degeneracies. 
Adding more features to the NN comes with a caveat: the more features fed to a NN the more training samples it needs to learn the mapping, thus setting more stringent requirements on the spectroscopic set. Although given that the size and ellipticity space is likely to be densely sampled it remains to be seen if this is an issue.

All the photometric redshifts for CFHTLenS  will be made  publicly available in the near future \footnote{send in inquiries to c.bonnett@gmail.com}. 
 
\section*{Acknowledgments}
We thank Hendrik Hildebrand, Catherine Heymans, Carles Sanchez  and Enrique Gazta\~naga  for helpful discussions related to this work.  
We thank Philip Graff for help with the $SkyNet$ software and useful comments related to this work. 

CB is supported by the Spanish Science MinistryAYA2009-13936 Consolider-Ingenio CSD2007-00060, project2009SGR1398 from Generalitat de Catalunya and by the the European CommissionÕs Marie Curie Initial Training Network CosmoComp (PITN-GA-2009-238356)

This work is based on observations obtained with MegaPrime/MegaCam, a joint project of CFHT and CEA/IRFU, at the Canada-France-Hawaii Telescope (CFHT) which is operated by the National Research Council (NRC) of Canada, the Institut National des Sciences de l'Univers of the Centre National de la Recherche Scientifique (CNRS) of France, and the University of Hawaii. This research used the facilities of the Canadian Astronomy Data Centre operated by the National Research Council of Canada with the support of the Canadian Space Agency. CFHTLenS data processing was made possible thanks to significant computing support from the NSERC Research Tools and Instruments grant program.

This paper uses data from the VIMOS Public Extragalactic Redshift Survey (VIPERS). VIPERS has been performed using the ESO Very Large Telescope, under the "Large Programme" 182.A-0886. The participating institutions and funding agencies are listed at http://vipers.inaf.it

This research uses data from the VIMOS VLT Deep Survey, obtained from the VVDS database operated by Cesam, Laboratoire d'Astrophysique de Marseille, France.

Funding for the DEEP2 Galaxy Redshift Survey has been
provided by NSF grants AST-95-09298, AST-0071048, AST-0507428, and AST-0507483 as well as NASA LTSA grant NNG04GC89G.

\bibliography{NN} % but remove .bib

\begin{thebibliography}{}
 \providecommand{\href}[2]{#2}
  \providecommand{\doi}[1]{\href{http://dx.doi.org/#1}{doi:#1}}
  \providecommand{\eprint}[1]{\href{http://arxiv.org/abs/#1}{arXiv:#1}}

\bibitem[\protect\citeauthoryear{MacKay}{MacKay}{2003}]{MacKay}
MacKay D. J.~C.,  2003, Information Theory, Inference, and Learning Algorithms.
Cambridge University Press

\bibitem[\protect\citeauthoryear{P\'erez \& Granger}{P\'erez \&
  Granger}{2007}]{ipython}
P\'erez F.,  Granger B.~E.,  2007, {C}omput. {S}ci. {E}ng., 9, 21

\bibitem[\protect\citeauthoryear{Schaul, Bayer, Wierstra, Sun, Felder, Sehnke,
  R{\"u}ckstie{\ss} \& Schmidhuber}{Schaul et~al.}{2010}]{pybrain}
Schaul T.,  Bayer J.,  Wierstra D.,  Sun Y.,  Felder M.,  Sehnke F.,
  R{\"u}ckstie{\ss} T.,    Schmidhuber J.,  2010, Journal of Machine Learning
  Research, 11, 743

\bibitem[\protect\citeauthoryear{{Albrecht} et~al.,}{{Albrecht}
  et~al.}{2006}]{DETF}
{Albrecht} A.  et~al., 2006, ArXiv Astrophysics e-prints,
  \eprint{astro-ph/0609591}

\bibitem[\protect\citeauthoryear{{Arnouts}, {Cristiani}, {Moscardini},
  {Matarrese}, {Lucchin}, {Fontana} \& {Giallongo}}{{Arnouts}
  et~al.}{1999}]{LEPHARE}
{Arnouts} S.,  {Cristiani} S.,  {Moscardini} L.,  {Matarrese} S.,  {Lucchin}
  F.,  {Fontana} A.,    {Giallongo} E.,  1999, Monthly Notices of the RAS, 310,
  540, \eprint{astro-ph/9902290}, \doi{10.1046/j.1365-8711.1999.02978.x}

\bibitem[\protect\citeauthoryear{{Bartelmann} \& {Schneider}}{{Bartelmann} \&
  {Schneider}}{2001}]{bs}
{Bartelmann} M.,  {Schneider} P.,  2001, Physics Reports, 340, 291,
  \eprint{astro-ph/9912508}, \doi{10.1016/S0370-1573(00)00082-X}

\bibitem[\protect\citeauthoryear{{Benjamin}, {van Waerbeke}, {M{\'e}nard} \&
  {Kilbinger}}{{Benjamin} et~al.}{2010}]{ben2010}
{Benjamin} J.,  {van Waerbeke} L.,  {M{\'e}nard} B.,    {Kilbinger} M.,  2010,
  Monthly Notices of the RAS, 408, 1168, \eprint{1002.2266},
  \doi{10.1111/j.1365-2966.2010.17191.x}

\bibitem[\protect\citeauthoryear{{Ben{\'{\i}}tez}}{{Ben{\'{\i}}tez}}{2011}]{BP%
Z}
{Ben{\'{\i}}tez} N.,  2011, {BPZ: Bayesian Photometric Redshift Code}

\bibitem[\protect\citeauthoryear{{Carrasco Kind} \& {Brunner}}{{Carrasco Kind}
  \& {Brunner}}{2013}]{carrasco}
{Carrasco Kind} M.,  {Brunner} R.~J.,  2013, Monthly Notices of the RAS, 432,
  1483, \eprint{1303.7269}, \doi{10.1093/mnras/stt574}

\bibitem[\protect\citeauthoryear{{Cavuoti}, {Brescia}, {Longo} \&
  {Mercurio}}{{Cavuoti} et~al.}{2012}]{c2012}
{Cavuoti} S.,  {Brescia} M.,  {Longo} G.,    {Mercurio} A.,  2012, Astronomy
  and Astrophysics, 546, A13, \eprint{1206.0876},
  \doi{10.1051/0004-6361/201219755}

\bibitem[\protect\citeauthoryear{{Collister} \& {Lahav}}{{Collister} \&
  {Lahav}}{2004}]{ANNZ}
{Collister} A.~A.,  {Lahav} O.,  2004, Publications of the ASP, 116, 345,
  \eprint{astro-ph/0311058}, \doi{10.1086/383254}

\bibitem[\protect\citeauthoryear{{Cunha}, {Lima}, {Oyaizu}, {Frieman} \&
  {Lin}}{{Cunha} et~al.}{2009}]{cuncha}
{Cunha} C.~E.,  {Lima} M.,  {Oyaizu} H.,  {Frieman} J.,    {Lin} H.,  2009,
  Monthly Notices of the RAS, 396, 2379, \eprint{0810.2991},
  \doi{10.1111/j.1365-2966.2009.14908.x}

\bibitem[\protect\citeauthoryear{{Cunha}, {Huterer}, {Lin}, {Busha} \&
  {Wechsler}}{{Cunha} et~al.}{2012a}]{cunhab}
{Cunha} C.~E.,  {Huterer} D.,  {Lin} H.,  {Busha} M.~T.,    {Wechsler} R.~H.,
  2012a, ArXiv e-prints, \eprint{1207.3347}

\bibitem[\protect\citeauthoryear{{Cunha}, {Huterer}, {Busha} \&
  {Wechsler}}{{Cunha} et~al.}{2012b}]{cunhaa}
{Cunha} C.~E.,  {Huterer} D.,  {Busha} M.~T.,    {Wechsler} R.~H.,  2012b,
  Monthly Notices of the RAS, 423, 909, \eprint{1109.5691},
  \doi{10.1111/j.1365-2966.2012.20927.x}

\bibitem[\protect\citeauthoryear{{Davis} et~al.,}{{Davis} et~al.}{2007}]{deep2}
{Davis} M.  et~al., 2007, Astrophysical Journal, Letters, 660, L1,
  \eprint{astro-ph/0607355}, \doi{10.1086/517931}

\bibitem[\protect\citeauthoryear{{Erben} et~al.,}{{Erben} et~al.}{2013}]{erben}
{Erben} T.  et~al., 2013, Monthly Notices of the RAS, 433, 2545,
  \eprint{1210.8156}, \doi{10.1093/mnras/stt928}

\bibitem[\protect\citeauthoryear{{Feldmann} et~al.,}{{Feldmann}
  et~al.}{2006}]{ZEBRA}
{Feldmann} R.  et~al., 2006, Monthly Notices of the RAS, 372, 565,
  \eprint{astro-ph/0609044}, \doi{10.1111/j.1365-2966.2006.10930.x}

\bibitem[\protect\citeauthoryear{{Freeman}, {Newman}, {Lee}, {Richards} \&
  {Schafer}}{{Freeman} et~al.}{2009}]{freeman2009}
{Freeman} P.~E.,  {Newman} J.~A.,  {Lee} A.~B.,  {Richards} J.~W.,    {Schafer}
  C.~M.,  2009, Monthly Notices of the RAS, 398, 2012, \eprint{0906.0995},
  \doi{10.1111/j.1365-2966.2009.15236.x}

\bibitem[\protect\citeauthoryear{{Garilli} et~al.,}{{Garilli}
  et~al.}{2008}]{vvdsf22}
{Garilli} B.  et~al., 2008, Astronomy and Astrophysics, 486, 683,
  \eprint{0804.4568}, \doi{10.1051/0004-6361:20078878}

\bibitem[\protect\citeauthoryear{{Garilli} et~al.,}{{Garilli}
  et~al.}{2013}]{vipers}
{Garilli} B.  et~al., 2013, ArXiv e-prints, \eprint{1310.1008}

\bibitem[\protect\citeauthoryear{{Geach}}{{Geach}}{2012}]{geach2012}
{Geach} J.~E.,  2012, Monthly Notices of the RAS, 419, 2633,
  \eprint{1110.0005}, \doi{10.1111/j.1365-2966.2011.19913.x}

\bibitem[\protect\citeauthoryear{{Gerdes}, {Sypniewski}, {McKay}, {Hao},
  {Weis}, {Wechsler} \& {Busha}}{{Gerdes} et~al.}{2010}]{AZ}
{Gerdes} D.~W.,  {Sypniewski} A.~J.,  {McKay} T.~A.,  {Hao} J.,  {Weis} M.~R.,
  {Wechsler} R.~H.,    {Busha} M.~T.,  2010, Astrophysical Journal, 715, 823,
  \eprint{0908.4085}, \doi{10.1088/0004-637X/715/2/823}

\bibitem[\protect\citeauthoryear{{Graff}, {Feroz}, {Hobson} \&
  {Lasenby}}{{Graff} et~al.}{2013}]{graff}
{Graff} P.,  {Feroz} F.,  {Hobson} M.~P.,    {Lasenby} A.~N.,  2013, ArXiv
  e-prints, \eprint{1309.0790}

\bibitem[\protect\citeauthoryear{{Heymans} et~al.,}{{Heymans}
  et~al.}{2012}]{heymans}
{Heymans} C.  et~al., 2012, Monthly Notices of the RAS, 427, 146,
  \eprint{1210.0032}, \doi{10.1111/j.1365-2966.2012.21952.x}

\bibitem[\protect\citeauthoryear{{Heymans} et~al.,}{{Heymans}
  et~al.}{2013}]{heymans2012}
{Heymans} C.  et~al., 2013, Monthly Notices of the RAS, 432, 2433,
  \eprint{1303.1808}, \doi{10.1093/mnras/stt601}

\bibitem[\protect\citeauthoryear{{Hildebrandt} et~al.,}{{Hildebrandt}
  et~al.}{2010}]{phat}
{Hildebrandt} H.  et~al., 2010, Astronomy and Astrophysics, 523, A31,
  \eprint{1008.0658}, \doi{10.1051/0004-6361/201014885}

\bibitem[\protect\citeauthoryear{{Hildebrandt} et~al.,}{{Hildebrandt}
  et~al.}{2012}]{HH}
{Hildebrandt} H.  et~al., 2012, Monthly Notices of the RAS, 421, 2355,
  \eprint{1111.4434}, \doi{10.1111/j.1365-2966.2012.20468.x}

\bibitem[\protect\citeauthoryear{{Jasche} \& {Wandelt}}{{Jasche} \&
  {Wandelt}}{2012}]{jj}
{Jasche} J.,  {Wandelt} B.~D.,  2012, Monthly Notices of the RAS, 425, 1042,
  \eprint{1106.2757}, \doi{10.1111/j.1365-2966.2012.21423.x}

\bibitem[\protect\citeauthoryear{{Kilbinger} et~al.,}{{Kilbinger}
  et~al.}{2009}]{nc}
{Kilbinger} M.  et~al., 2009, Astronomy and Astrophysics, 497, 677,
  \eprint{0810.5129}, \doi{10.1051/0004-6361/200811247}

\bibitem[\protect\citeauthoryear{{Le F{\`e}vre} et~al.,}{{Le F{\`e}vre}
  et~al.}{2005}]{vvds}
{Le F{\`e}vre} O.  et~al., 2005, Astronomy and Astrophysics, 439, 877,
  \eprint{astro-ph/0409135}, \doi{10.1051/0004-6361:20041962}

\bibitem[\protect\citeauthoryear{{McQuinn} \& {White}}{{McQuinn} \&
  {White}}{2013}]{white2013}
{McQuinn} M.,  {White} M.,  2013, Monthly Notices of the RAS, 433, 2857,
  \eprint{1302.0857}, \doi{10.1093/mnras/stt914}

\bibitem[\protect\citeauthoryear{{M{\'e}nard}, {Scranton}, {Schmidt},
  {Morrison}, {Jeong}, {Budavari} \& {Rahman}}{{M{\'e}nard}
  et~al.}{2013}]{menard}
{M{\'e}nard} B.,  {Scranton} R.,  {Schmidt} S.,  {Morrison} C.,  {Jeong} D.,
  {Budavari} T.,    {Rahman} M.,  2013, ArXiv e-prints, \eprint{1303.4722}

\bibitem[\protect\citeauthoryear{{Newman}}{{Newman}}{2008}]{new2008}
{Newman} J.~A.,  2008, Astrophysical Journal, 684, 88, \eprint{0805.1409},
  \doi{10.1086/589982}

\bibitem[\protect\citeauthoryear{{Newman} et~al.,}{{Newman}
  et~al.}{2013}]{new2013}
{Newman} J.  et~al., 2013, ArXiv e-prints, \eprint{1309.5384}

\bibitem[\protect\citeauthoryear{{Smith} et~al.,}{{Smith}
  et~al.}{2003}]{sm2003}
{Smith} R.~E.  et~al., 2003, Monthly Notices of the RAS, 341, 1311,
  \eprint{astro-ph/0207664}, \doi{10.1046/j.1365-8711.2003.06503.x}

\bibitem[\protect\citeauthoryear{{Vanzella} et~al.,}{{Vanzella}
  et~al.}{2004}]{vv}
{Vanzella} E.  et~al., 2004, Astronomy and Astrophysics, 423, 761,
  \eprint{astro-ph/0312064}, \doi{10.1051/0004-6361:20040176}

\bibitem[\protect\citeauthoryear{{Wadadekar}}{{Wadadekar}}{2005}]{svm2005}
{Wadadekar} Y.,  2005, Publications of the ASP, 117, 79,
  \eprint{astro-ph/0412005}, \doi{10.1086/427710}

\bibitem[\protect\citeauthoryear{{Way}, {Foster}, {Gazis} \&
  {Srivastava}}{{Way} et~al.}{2009}]{way2009}
{Way} M.~J.,  {Foster} L.~V.,  {Gazis} P.~R.,    {Srivastava} A.~N.,  2009,
  Astrophysical Journal, 706, 623, \eprint{0905.4081},
  \doi{10.1088/0004-637X/706/1/623}

\bibitem[\protect\citeauthoryear{{Wolf}}{{Wolf}}{2009}]{wolf}
{Wolf} C.,  2009, Monthly Notices of the RAS, 397, 520, \eprint{0904.3438},
  \doi{10.1111/j.1365-2966.2009.14953.x}

\end{thebibliography}

\bsp

%  $N(z)_{NN}^{truth}$ and $N(z)_{NN}$
\appendix

\begin{table}
\center
\begin{tabular}{l*{4}{c}r}
z-bin   & $D_1$    & $D_2$ & $D_3$ & $D_4$  \\
\hline
all   z &  0.05 & 0.0109   &  0.0028  & 0.067    \\
bin 1 &  0.18  &  0.0712  &  0.0115   & 0.37   \\
bin 2 &  0.07  &  0.0324  &  0.0062   & 0.20 \\
bin 3 &  0.09  &  0.0180  &  0.0032   &  0.17  \\
bin 4 &  0.08  &  0.0291  &  0.0050   &  0.12 \\
bin 5 &  0.04  & 0.0260   &  0.0038  &   0.10  \\
bin 6 &  0.07  & 0.0303   &  0.0079   &   0.08 \\
bin 7 &  0.07  & 0.0553   &  0.0149   &   0.11 \\
bin 8 &  0.19   & 0.1053   &  0.0135   &  0.42 \\ 
\end{tabular}
\caption{The same as table \ref{D_table} with the $D_4$ values  added. The  $D_4$ values are the mean $D$ from  the 50 validation  samples for the redshifts  derived using $BPZ$  in \citep{HH} .
The $D_1$ values are those from the test set (see Sect.\ref{bla} ). $D_2$ is the maximum $D$ value of the 50 validation samples when trained on $70\%$ of the galaxies. $D_2$ is thus the worst result for each redshift bin from Sect. \ref{qwert}.
$D_3$ is the mean $D$ from the 50 validation samples in Sect. \ref{qwert} .
All 4 cases use 5 band photometry information.}
\label{D2_table}
\end{table}

\section{Comparison with BPZ} \label{app_bpz}
\begin{figure*}
\centering
\includegraphics[scale=0.30]{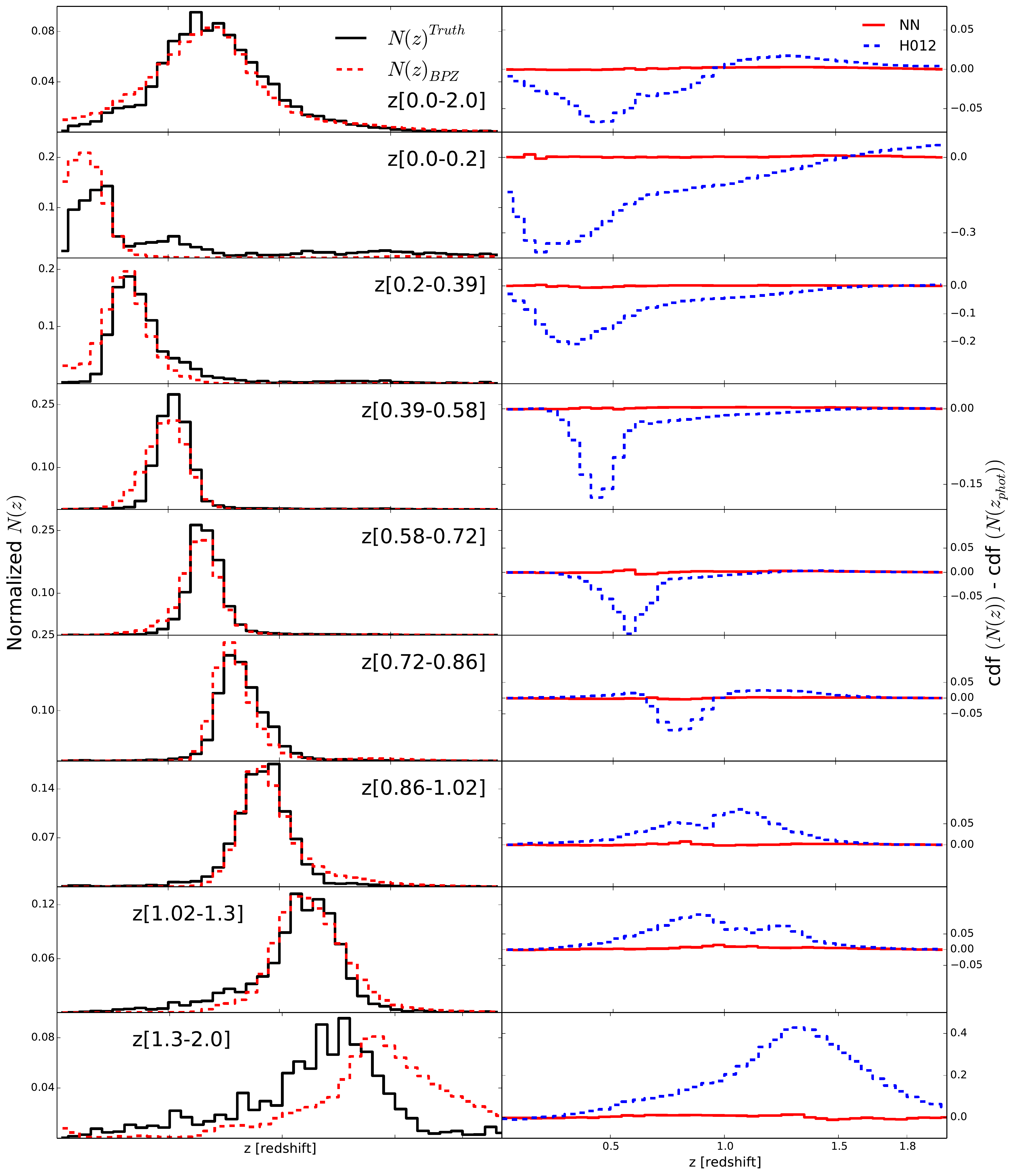}
\caption{The left column shows the  mean $N(z)_{PBZ}^{truth}$ (solid black line) and $N(z)_{BPZ}$ (dotted red line) of the 50 validation sets for the full distribution and the 8  tomographic  bins.
The right column shows the  mean difference in the the cumulative function $N(z_{phot})$  and $N(z)$ for both BPZ (dotted line) and the NN (solid line). 
Is it clear for NN performs significantly better than BPZ on the validation set.} 
\label{bpz_all}
\end{figure*}

This appendix compares the redshifts obtained by H012, using the BPZ \footnote{http://acs.pha.jhu.edu/~txitxo/} code  and this work. 
In  the left  column in Figure \ref{bpz_all} we show the mean $N(z)_{PBZ}^{truth}$ and $N(z)_{BPZ}$ of the 50 validation sets (17817 galaxies per set) in 9 redshift bins.
We use the $z_{phot}$ as returned by $BPZ$ to select galaxies, this coincides  with the mode of the PDF returned  by the code.   
The PDF returned for the galaxies has the same $z$ resolution as the NN output ($\Delta z =0.05$).
The $N(z)^{truth}$ is obtained by binning the spectra of the selected galaxies with the same $z$ resolution.
The right column in Figure \ref{bpz_all} shows the difference of the cumulative distributions for the results obtained in Sect. \ref{qwert} and those of H012. 
Table \ref{D2_table} shows a summary of  the $D$ statistics (Eq. \ref{ds}) obtained in this study. It is the same as table \ref{D_table} but the $D$ (as $D_4$) values for the photometric redshifts from H012 have been added. 
The $D_4$  values are of the same order as the $D$ values that where obtained in Sect. \ref{bla}, where the test set was not represented in the training set.

\section{Transition matrix}\label{app_trans}
This appendix shows the transition  matrices (\ref{tm5}, \ref{tm4} and \ref{tm3}) for the 8 redshift bins studied in Sect. \ref{qwert}.
The transition matrix shows the percentage of the galaxies that are in each tomographic bin for a given redshift selection.
The diagonal thus shows the fraction of galaxies that lie in the tomographic bin on which the galaxies where selected. 
The off diagonal terms show the fraction leaked to the other bins.  
We show two rows for each bin. The top row is the predicted transition estimated by the NN, the bottom row is the actual transition.
The errors are the standard deviation form the 50 validation samples.  
In left most column the average amount of galaxies selected in each bin is given.
For the selection in tomographic bin 3, 4, 5 and 7 the transition matrices show that the NN overestimates the amount of objects in bin 8 using 5 bands.
When using 4 or 3 bands this only happens in tomographic bin 3 and 4.

\begin{table*} 
\begin{tabular}{ |l| l| l | l |l | l| l | l| l|}
\hline
 \multicolumn{9}{ |c| }{CFHTLenS 5 bands} \\
 \hline
 \#gal &   bin1 & bin 2  & bin3 & bin 4 & bin 5 & bin 6 & bin 7 & bin 8   \\
&  z[0 -0.2] & z[0.2-0.39] & z[0.39-0.58]    & z[0.58-0.72]   & z[0.72-0.86]   & z[0.86-1.02]  & z[1.02-1.3]  & z[1.3-2.0] \\
  \hline 
371     & 0.59   $\pm$ 0.03 & 0.14 $\pm$ 0.01 & 0.10 $\pm$0.01 & 0.024 $\pm$  0 .004 & 0.015 $\pm$ 0.003  & 0.02  $\pm$ 0.01 & 0.03 $\pm$ 0.01 &  0.07 $\pm$ 0.01 \\
 bin 1         & 0.59  $\pm$ 0.04 & 0.15 $\pm$ 0.02 & 0.10 $\pm$0.02 & 0.025 $\pm$  0 .007 & 0.018 $\pm$ 0.006  & 0.02  $\pm$ 0.01 & 0.03 $\pm$ 0.01 &  0.07 $\pm$ 0.01   \\
\hline
1374 & 0.054  $\pm$ 0.03  &  0.70 $\pm$ 0.02 & 0.136 $\pm $ 0.008 & 0.026 $\pm$ 0.002 & 0.017 $ \pm$0.002  &  0.008 $\pm$0.002 & 0.022 $\pm$ 0.003 & 0.033 $\pm$ 0.04 \\
 bin 2     & 0.057  $\pm$ 0.06  &  0.69 $\pm$ 0.02  & 0.142 $\pm$ 0.011 &  0.026 $\pm$ 0.004 &  0.019$ \pm$ 0.003 & 0.007 $\pm$ 0.002 & 0.023 $\pm$0.003 & 0.032 $\pm$ 0.04 \\
 \hline
4041& 0.007  $\pm$ 0.001  &  0.055 $\pm$ 0.003  &  0.78 $\pm$ 0.01 &  0.104  $\pm$ 0.003 &  0.014  $\pm$ 0.002 & 0.007 $\pm$ 0.001 & 0.013  $\pm$  0.002  &  0.018 $\pm$ 0.002 \\
  bin 3                  & 0.007 $\pm$ 0.001  &  0.057 $\pm$ 0.003 &  0.78 $\pm$0.01 &  0.106$\pm$  0.005  &  0.015  $\pm$ 0.002 & 0.007 $\pm$ 0.001 & 0.012  $\pm$   0.002 &  0.014 $\pm$ 0.002\\
 \hline
3301 & 0.003  $\pm$ 0.001  &  0.011 $\pm$ 0.001  &  0.193 $\pm$ 0.005 &  0.60  $\pm$  0.01  &  0.135  $\pm$ 0.006 & 0.018 $\pm$ 0.001 & 0.021  $\pm$  0.002  &  0.013 $\pm$ 0.001  \\
 bin 4   & 0.002 $\pm$  0.001  &  0.011 $\pm$ 0.002  &  0.198 $\pm$ 0.008 &  0.59   $\pm$  0.01 &  0.140  $\pm$ 0.008 & 0.019 $\pm$ 0.002 & 0.022  $\pm$  0.003  &  0.010 $\pm$ 0.002 \\
 \hline
4376& 0.0029  $\pm$ 0.0009  &  0.005 $\pm$ 0.001  &  0.023 $\pm$ 0.002 &  0.119  $\pm$  0.003  &  0.62  $\pm$ 0.01 & 0.167 $\pm$ 0.007  & 0.046  $\pm$  0.002  &  0.0138 $\pm$ 0.002\\
 bin 5               & 0.0020  $\pm$ 0.0006  &  0.004 $\pm$ 0.001  &  0.024 $\pm$ 0.002 &  0.122  $\pm$  0.005  &  0.62  $\pm$ 0.01 & 0.172 $\pm$ 0.008 & 0.047  $\pm$  0.003  &   0.0116 $\pm$ 0.002 \\
\hline
2585 & 0.005  $\pm$ 0.001  &  0.005 $\pm$ 0.001  &  0.019 $\pm$ 0.002  &  0.023  $\pm$  0.002  &    0.24 $\pm$ 0.001    & 0.53  $\pm$ 0.02 & 0.158 $\pm$ 0.008    & 0.021  $\pm$  0.003   \\
  bin 6               & 0.005  $\pm$ 0.001  &  0.004 $\pm$ 0.001   &  0.020 $\pm$ 0.003 &  0.025   $\pm$  0.003  &    0.24     $\pm$ 0.001 &  0.52  $\pm$ 0.01 & 0.160 $\pm$ 0.011  & 0.020  $\pm$  0.003   \\
\hline
1348      & 0.006  $\pm$ 0.001  &  0.014 $\pm$ 0.002  &  0.031 $\pm$ 0.003  &  0.032  $\pm$  0.003   &    0.101 $\pm$ 0.006    & 0.142  $\pm$ 0.009 & 0.57 $\pm$ 0.02    & 0.130  $\pm$  0.006\\
  bin 7     & 0.006  $\pm$ 0.002  &  0.017 $\pm$ 0.003  &  0.031 $\pm$ 0.004  &  0.043  $\pm$  0.005   &    0.107 $\pm$ 0.010    & 0.149  $\pm$ 0.012 & 0.56 $\pm$ 0.02    & 0.098  $\pm$  0.009 \\
\hline
216      &  0.033  $\pm$ 0.006  &  0.050 $\pm$ 0.008  &  0.061 $\pm$ 0.009  &  0.024  $\pm$  0.005   &    0.038 $\pm$ 0.008   &  0.036  $\pm$ 0.007 & 0.23 $\pm$ 0.03    & 0.52  $\pm$  0.05\\
bin 8   &  0.030  $\pm$  0.011  &  0.054 $\pm$ 0.012  &  0.069 $\pm$ 0.016  &  0.027  $\pm$  0.008   &    0.037 $\pm$ 0.014   &  0.035  $\pm$ 0.012 & 0.24 $\pm$ 0.04    & 0.51  $\pm$  0.05 \\
 \hline
\end{tabular}
\caption{5 band transition matrix. Note that for bins 3, 4, 5 and 7 the NN over estimates the amount of galaxies in bin 8.}
\label{tm5}
\end{table*}

\begin{table*} 
\begin{tabular}{ |l| l| l | l |l | l| l | l| l|}
\hline
 \multicolumn{9}{ |c| }{CFHTLenS 4 bands} \\
 \hline
 \#gal &   bin1 & bin 2  & bin3 & bin 4 & bin 5 & bin 6 & bin 7 & bin 8   \\
&  z[0 -0.2] & z[0.2-0.39] & z[0.39-0.58]    & z[0.58-0.72]   & z[0.72-0.86]   & z[0.86-1.02]  & z[1.02-1.3]  & z[1.3-2.0] \\
  \hline 
284             & 0.55   $\pm$ 0.04 & 0.14 $\pm$ 0.01 & 0.09 $\pm$0.01 & 0.036   $\pm$  0 .004 & 0.021 $\pm$ 0.004  & 0.011  $\pm$ 0.003 & 0.043 $\pm$ 0.008 &  0.10 $\pm$ 0.016 \\
 bin 1         & 0.55   $\pm$ 0.04 & 0.14 $\pm$ 0.02 & 0.10 $\pm$0.01 & 0.037   $\pm$  0 .010 & 0.022 $\pm$ 0.009  & 0.013  $\pm$ 0.006 & 0.037 $\pm$ 0.011 &  0.10 $\pm$ 0.018   \\
\hline
1164       & 0.083  $\pm$ 0.006  &  0.60 $\pm$ 0.02 & 0.208 $\pm $ 0.008 &  0.026 $\pm$ 0.002 & 0.012 $ \pm$ 0.002 &  0.006  $\pm$ 0.001 & 0.024 $\pm$ 0.003 & 0.031 $\pm$ 0.003 \\
 bin 2      & 0.083  $\pm$ 0.008 &  0.60 $\pm$ 0.02  & 0.209 $\pm$ 0.015  &  0.028 $\pm$ 0.004 &  0.014$ \pm$ 0.003 &  0.005  $\pm$ 0.002 & 0.026 $\pm$ 0.005 & 0.030 $\pm$ 0.004 \\
\hline
4189           & 0.009  $\pm$ 0.001  &  0.105 $\pm$ 0.003  &  0.73 $\pm$ 0.01 &  0.103  $\pm$ 0.003 &  0.014  $\pm$ 0.002 & 0.007 $\pm$ 0.001 & 0.010  $\pm$  0.001  &  0.016 $\pm$ 0.002 \\
  bin 3         & 0.009 $\pm$ 0.001  &  0.105  $\pm$ 0.006 &  0.73 $\pm$0.01  &  0.104$\pm$  0.005  &  0.016  $\pm$ 0.002 & 0.007 $\pm$ 0.001 & 0.011  $\pm$   0.002 &  0.013 $\pm$ 0.001\\
\hline
3414        & 0.008  $\pm$ 0.001  &  0.021 $\pm$ 0.002  &  0.195 $\pm$ 0.008 &  0.58  $\pm$  0.01  &  0.137  $\pm$ 0.005 & 0.020 $\pm$ 0.002 & 0.023  $\pm$  0.002  &  0.014 $\pm$ 0.002  \\
 bin 4       & 0.007 $\pm$  0.002  &  0.021 $\pm$ 0.003  &  0.202 $\pm$ 0.010 &  0.57   $\pm$  0.01 &  0.143  $\pm$ 0.008 & 0.022 $\pm$ 0.003 & 0.023  $\pm$  0.003  &  0.010 $\pm$ 0.002 \\
\hline
4407        & 0.0039  $\pm$ 0.0010  &  0.006 $\pm$ 0.0010  &  0.023 $\pm$ 0.002 &  0.119  $\pm$  0.004  &  0.61  $\pm$ 0.01 & 0.172 $\pm$ 0.006  & 0.048  $\pm$  0.004  &   0.0131 $\pm$ 0.002\\
 bin 5       & 0.0031  $\pm$ 0.0008  &  0.005 $\pm$ 0.0008 &  0.025 $\pm$ 0.002  &  0.122  $\pm$  0.006  &  0.61  $\pm$ 0.01 & 0.175 $\pm$ 0.009  & 0.050  $\pm$  0.005  &   0.0114 $\pm$ 0.002 \\
\hline
2555       & 0.0048  $\pm$ 0.001  &  0.0048 $\pm$ 0.0010  &  0.019 $\pm$ 0.002  &  0.026  $\pm$  0.003   &    0.24 $\pm$ 0.01   & 0.52  $\pm$ 0.02  & 0.16 $\pm$ 0.01   & 0.023  $\pm$  0.003\\
 bin 6     & 0.0049  $\pm$ 0.002  &  0.0041 $\pm$ 0.0011  &  0.020 $\pm$ 0.002  &  0.027  $\pm$  0.003   &     0.24 $\pm$ 0.01   & 0.52  $\pm$ 0.01  & 0.16 $\pm$ 0.01   & 0.023 $\pm$  0.003 \\
\hline
1438          & 0.011  $\pm$ 0.002  &  0.02 $\pm$ 0.002   &  0.036 $\pm$ 0.003  &  0.034  $\pm$  0.003   &    0.10     $\pm$ 0.007    & 0.14  $\pm$ 0.01 & 0.54 $\pm$ 0.02    & 0.12  $\pm$  0.01   \\
 bin 7        & 0.011  $\pm$ 0.003  &  0.02 $\pm$ 0.003   &  0.037 $\pm$ 0.004  &  0.033   $\pm$  0.004  &    0.11     $\pm$ 0.009  &  0.14  $\pm$  0.01 & 0.53 $\pm$ 0.02    & 0.12  $\pm$  0.01   \\
\hline
163       &  0.06  $\pm$  0.01  &  0.06 $\pm$ 0.01  &  0.058 $\pm$ 0.01  &  0.021  $\pm$  0.005   &    0.039 $\pm$ 0.009   &  0.033  $\pm$ 0.008 &  0.24 $\pm$ 0.03    & 0.49  $\pm$  0.06\\
bin 8      &  0.06  $\pm$  0.01  &  0.06 $\pm$ 0.02  &  0.057 $\pm$ 0.02  &  0.023  $\pm$  0.012   &    0.038 $\pm$ 0.017   &  0.033  $\pm$ 0.009 &  0.24 $\pm$ 0.04    & 0.48  $\pm$  0.06 \\
 \hline
\end{tabular}
\caption{4 band transition matrix. Note that for bins 3 and 4 the NN over estimates the amount of galaxies in bin 8. }
\label{tm4}
\end{table*}

\begin{table*} 
\begin{tabular}{ |l| l| l | l |l | l| l | l| l|}
\hline
 \multicolumn{9}{ |c| }{CFHTLenS 3 bands} \\
 \hline
 \#gal &   bin1 & bin 2  & bin3 & bin 4 & bin 5 & bin 6 & bin 7 & bin 8   \\
&  z[0 -0.2] & z[0.2-0.39] & z[0.39-0.58]    & z[0.58-0.72]   & z[0.72-0.86]   & z[0.86-1.02]  & z[1.02-1.3]  & z[1.3-2.0] \\
 \hline 
312             & 0.49  $\pm$ 0.04 & 0.14 $\pm$ 0.01  & 0.11 $\pm$0.01 & 0.05 $\pm$  0 .01 & 0.026 $\pm$ 0.005  & 0.018  $\pm$ 0.004 & 0.06 $\pm$ 0.01 &  0.10 $\pm$ 0.01 \\
 bin 1         & 0.50  $\pm$ 0.04 & 0.14 $\pm$ 0.03  & 0.11 $\pm$0.02 & 0.05 $\pm$  0 .01 & 0.030 $\pm$ 0.010  & 0.018  $\pm$ 0.006 & 0.05 $\pm$ 0.01 &  0.10 $\pm$ 0.02   \\
\hline
847        & 0.10  $\pm$ 0.006  &  0.53 $\pm$ 0.03  &  0.24 $\pm $ 0.02 & 0.070 $\pm$ 0.005 & 0.011  $ \pm$ 0.002  &  0.005 $\pm$0.001 & 0.018 $\pm$ 0.003 & 0.023 $\pm$ 0.003 \\
 bin 2     & 0.10 $\pm$ 0.011  &  0.52 $\pm$ 0.03  & 0.24 $\pm $ 0.03  &  0.076 $\pm$ 0.010 &  0.012$ \pm$ 0.003 &  0.002 $\pm$ 0.002 & 0.017 $\pm$0.004 & 0.021 $\pm$ 0.004 \\
 \hline
3982            & 0.007  $\pm$ 0.001  &  0.12 $\pm$ 0.005  &  0.72 $\pm$ 0.01 &  0.115  $\pm$ 0.003  &  0.014  $\pm$ 0.001 & 0.007 $\pm$ 0.001 & 0.009  $\pm$  0.001  &  0.014 $\pm$ 0.001 \\
  bin 3          & 0.007 $\pm$ 0.001   &  0.12 $\pm$ 0.005  &  0.71 $\pm$0.01 &   0.118 $\pm$  0.005  &  0.016  $\pm$ 0.002 & 0.006 $\pm$ 0.001 & 0.009  $\pm$   0.002 &  0.012 $\pm$ 0.001\\
\hline
3745    & 0.011  $\pm$ 0.001  &  0.07 $\pm$ 0.003  &  0.238 $\pm$ 0.007 &  0.491   $\pm$  0.009  &  0.138  $\pm$ 0.004 & 0.016 $\pm$ 0.001 & 0.016  $\pm$  0.002  &  0.016 $\pm$ 0.002  \\
 bin 4   & 0.010 $\pm$  0.002  &  0.07 $\pm$ 0.004  &  0.242 $\pm$ 0.008 &  0.486   $\pm$  0.009  &  0.14 3 $\pm$ 0.008 & 0.016 $\pm$ 0.002 & 0.016  $\pm$  0.002  &  0.013 $\pm$ 0.001 \\
\hline
5363          & 0.0062   $\pm$ 0.0010  &  0.011 $\pm$ 0.001  &  0.031 $\pm$ 0.002 &  0.119  $\pm$  0.003  &  0.51  $\pm$ 0.01 & 0.189 $\pm$ 0.006  &  0.104  $\pm$  0.004  &  0.023 $\pm$ 0.001\\
 bin 5         & 0.0052  $\pm$ 0.0008   &  0.011 $\pm$ 0.001  &  0.032 $\pm$ 0.002 &  0.123  $\pm$  0.005  &  0.51  $\pm$ 0.01 & 0.192 $\pm$ 0.009  & 0.105  $\pm$  0.005  &   0.021 $\pm$ 0.002 \\
\hline
2839          &  0.006  $\pm$ 0.001  &  0.008 $\pm$ 0.001   &  0.020 $\pm$ 0.002  &  0.023  $\pm$  0.002  &    0.216   $\pm$ 0.008    &  0.43  $\pm$ 0.01 & 0.25 $\pm$ 0.01  & 0.044  $\pm$  0.004   \\
bin 6          & 0.005   $\pm$ 0.001  &  0.009 $\pm$ 0.001   &  0.021 $\pm$ 0.002  &  0.024   $\pm$  0.003  &    0.223   $\pm$ 0.011   &  0.42  $\pm$ 0.01 & 0.25 $\pm$ 0.01  & 0.044  $\pm$  0.004   \\
\hline
370        & 0.018  $\pm$ 0.005  &  0.017 $\pm$ 0.004  &  0.033 $\pm$ 0.006  &  0.021  $\pm$  0.004   &    0.09 $\pm$ 0.01    & 0.20  $\pm$ 0.03  & 0.44 $\pm$ 0.05    & 0.17  $\pm$  0.03\\
bin 7     & 0.023  $\pm$ 0.006  &  0.017 $\pm$ 0.006  &  0.035 $\pm$ 0.009   &  0.015  $\pm$  0.006   &    0.11 $\pm$ 0.01    & 0.21  $\pm$ 0.04  & 0.42 $\pm$ 0.05    & 0.18  $\pm$  0.03 \\
\hline
154      &  0.06  $\pm$ 0.01   &  0.041 $\pm$ 0.008  &  0.07 $\pm$ 0.01  &  0.023  $\pm$  0.005   &    0.039 $\pm$ 0.009   &  0.08  $\pm$ 0.01 &  0.25  $\pm$ 0.05    & 0.43  $\pm$  0.07\\
bin 8   &  0.06  $\pm$  0.01  &  0.045 $\pm$ 0.015  &  0.07 $\pm$ 0.02  &  0.025 $\pm$   0.013   &    0.040 $\pm$ 0.015   &  0.09  $\pm$ 0.01 &  0.23  $\pm$ 0.06    & 0.44  $\pm$  0.06 \\
\hline
\end{tabular}
\caption{3 band transition matrix. Note that for bins 3 and 4 the NN over estimates the amount of galaxies in bin 8. }
\label{tm3}
\end{table*}

%\newpage
\section{Fitting function to the fractional error on $P{\kappa}$}
\begin{figure*}
\centering
\includegraphics[scale=0.9]{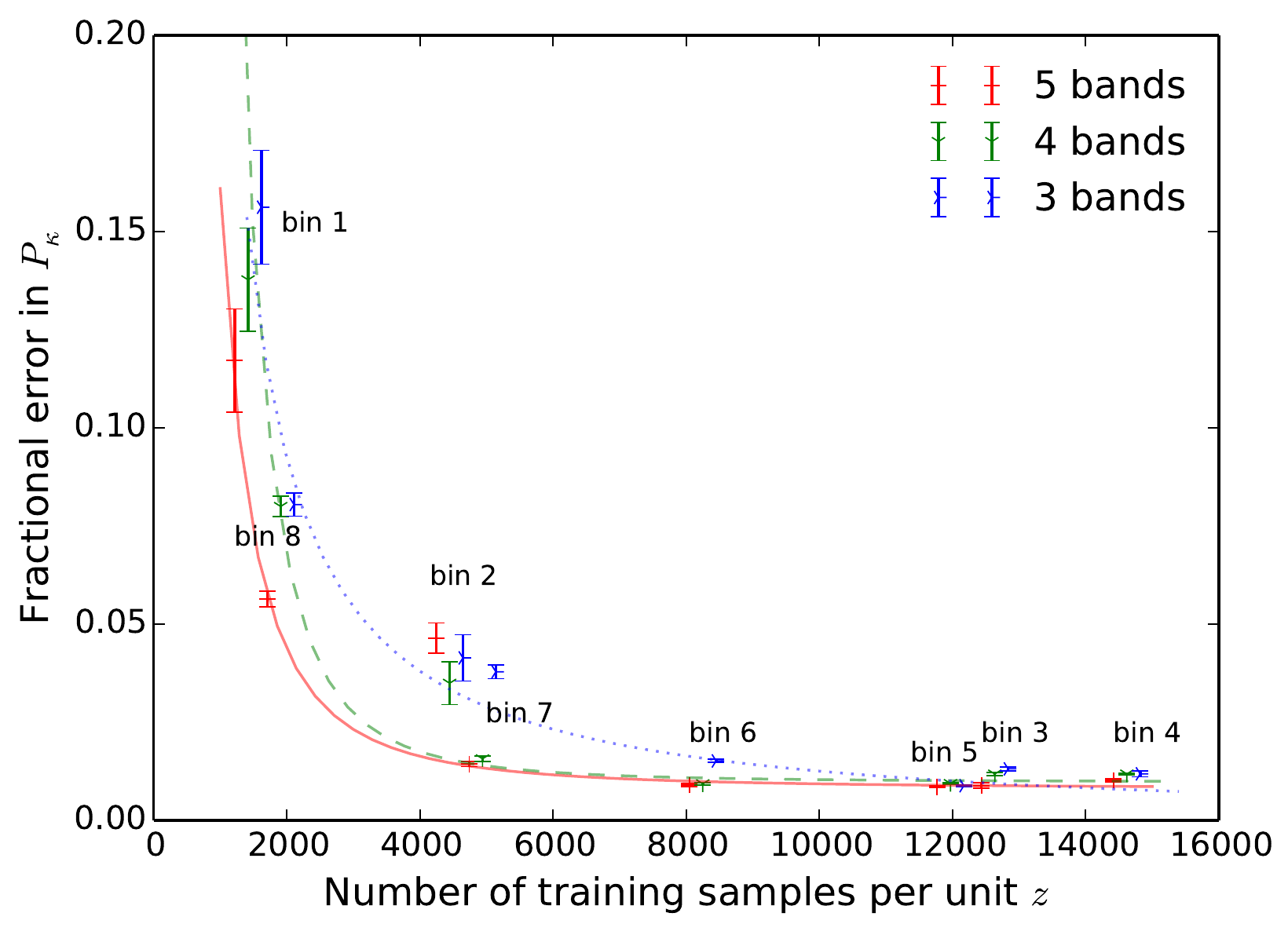}
\caption{This figure shows the fractional error on $P_{\kappa}$ as a function  of available training samples in the tomographic bin per unit redshift.
The error bars are give by the standard deviation of the error bar size over the range of $50 \leq l \leq 5000$.
The red solid line depicts the best fit for 5 bands, the green dashed line for 4 bands and the dotted blue line for 3 bands.
Tomographic bin 2 is outlier for the 5 and 4 band case. } 
\label{fit}
\end{figure*}

In this appendix we provide fitting functions to the fractional error on $P_{\kappa}$ as a function of the density of spectral redshift available in the tomographic bin.
We note that tomographic bin 2 is an outlier for the case of 5 and 4 bands and that the fitting function does not  provide a correct estimate for this bin, see Figure \ref{fit}.
$x$ is the amount of  training samples (70 \% of the full sample) available in the tomographic bin per unit redshift.
\begin{equation}
\begin{array}{l}
\displaystyle  P_{\kappa} error_{ugriz} = 3.26 \times 10 ^{5} \times x^{-2.11} + 8.15 \times 10^{-3} \\
\displaystyle  P_{\kappa} error_{griz}   = 3.33 \times 10 ^{7} \times x^{-2.69} + 9.81 \times 10^{-3}\\
\displaystyle  P_{\kappa} error_{grz}    = 2.59 \times 10 ^{2} \times x^{-1.07}  -1.03  \times 10^{-3}
\end{array} 
\label{eq:xdef}
\end{equation}
These fitting function are a first order approximation as they just look at the number of available training galaxies and do no take into account degeneracies between bins.       

\label{lastpage}
 
\end{document}